\pdfoutput=1

\documentclass{svproc}
\usepackage{hyperref}
\usepackage{pdfpages}
\usepackage{url}
\usepackage{amsmath}    % Math Library

\usepackage{hhline}			% Table

\usepackage{array}         % Align Package

\usepackage{fixltx2e}      % float package
\usepackage{dblfloatfix}
\usepackage{xcolor}
\usepackage{graphicx}
\usepackage{booktabs}
\usepackage{amsfonts}
\usepackage{chapterbib}

\newcommand{\PURPLE}[1]{{\textcolor{black}{#1}}} % 

\newcommand{\BLUE}[1]{{\textcolor{black}{#1}}} % 
\newcommand{\ORANGE}[1]{{\textcolor{black}{#1}}} % 
\newcommand{\RN}[1]{%
  \textup{\lowercase\expandafter{\romannumeral#1}}%
}

\allowdisplaybreaks

\begin{document}
\mainmatter              % start of a contribution
\title{Index Tracking via Learning to Predict Market Sensitivities}
\titlerunning{Index Tracking via Learning to Predict Market Sensitivities}  % abbreviated title (for running head)
\author{Yoonsik Hong, Yanghoon Kim, Jeonghun Kim, Yongmin Choi}
\authorrunning{Hong et al.} % abbreviated author list (for running head)
\tocauthor{Authors: Blind Peer Review}
\institute{
Mirae Asset Global Investments Co.,Ltd.\\
\email{ 
ys.hong@miraeasset.com, yanghoon.kim@miraeasset.com, jeonghun\_kim@miraeasset.com, yongmin.choi@miraeasset.com}
%Email: Blind Peer Review
% WWW home page:
% \texttt{http://users/\homedir iekeland/web/welcome.html}
% \and
% Universit\'{e} de Paris-Sud,
% Laboratoire d'Analyse Num\'{e}rique, B\^{a}timent 425,\\
% F-91405 Orsay Cedex, France
}

\maketitle              % typeset the title of the contribution

\begin{abstract}
%%%%%%%%%%%%%%%%%%%%%%%%%Intsruction%%%%%%%%%%%%%%%%%%%%%%%%%
% The abstract should summarize the contents of the paper
% using at least 150 and at most 200 words. It will be set in 9-point
% font size and be inset 1.0 cm from the right and left margins.
% There will be two blank lines before and after the Abstract. \dots
%%%%%%%%%%%%%%%%%%%%%%%%%Intsruction%%%%%%%%%%%%%%%%%%%%%%%%%
% 일반론
% however, 문제
% 우리는 해결한다
% 알고리즘/내용 간단 설명
% 우리꺼 좋다
% 하고 싶은 말- 우리꺼 정말 좋다

%%%%%%%%%%%%%%%%%%%%%%%%%%%%%%%%%%%%%%%%%%% 정리 %%%%%%%%%%%%%%%%%%%%%%%%%%%%%%%%%%%%%%%%%%%
% 일반론
% A significant number of equity funds are preferred by index funds nowadays, and market sensitivities are instrumental in managing them.
% \PURPLE{A significant number of index funds are preferred by investors nowadays, and market sensitivities are instrumental in managing index funds.} 
\BLUE{Index funds are substantially preferred by investors nowadays, and market sensitivities are instrumental in managing index funds.} 
\BLUE{An index fund is a mutual fund aiming to track the returns of a predefined market index (e.g., the S\&P 500).}
% however, 문제
% An index fund might replicate a market index identically, which is, however, cost-ineffective and impractical. 
\BLUE{A basic strategy to manage an index fund is replicating the index's constituents and weights identically, which is, however, cost-ineffective and impractical.} 
% \BLUE{Moreover, to utilize market sensitivities to replicate the index partially, they must be predicted or estimated accurately.}
% \RED{To address this issue, it is required to utilize market sensitivities to replicate the index partially and they must be predicted or estimated accurately.}
% \ORANGE{To address this issue, it is required to utilize accurately predicted market sensitivities to replicate the index partially.}
\ORANGE{To address this issue, it is required to replicate the index partially with accurately predicted market sensitivities.}
% 우리는 해결한다; 알고리즘/내용 간단 설명
% Accordingly, we propose a novel method that partially replicates the index via learning to predict market sensitivities.
\ORANGE{Accordingly, we propose a novel partial-replication method via learning to predict market sensitivities.}
% Accordingly, we propose a deep-learning framework for market sensitivities and mathematical programming for partial replication utilizing the sensitivities.
% Accordingly, first, we examine deep-learning models to predict market sensitivities. 
% First, we examine deep-learning models to predict market sensitivities. 
\ORANGE{We first examine deep-learning models to predict market sensitivities in a supervised manner with our data-processing methods.}
% Also \RED{(In Addition)}, we present pragmatic applications of data processing methods to aid training and generate target data for the prediction. 
Then, we propose a partial-index-tracking optimization model controlling the net predicted market sensitivities of the portfolios and index to be the same. 
These processes' efficacy is corroborated by \ORANGE{our experiments} on the Korea Stock Price Index 200. 
% 우리꺼 좋다
Our experiments show a significant reduction of the prediction errors compared with historical estimations and competitive tracking errors of replicating the index utilizing fewer than half of the entire constituents. 
% 하고 싶은 말- 우리꺼 정말 좋다
Therefore, we show that applying deep learning to predict market sensitivities is promising and that our portfolio construction methods are practically effective. Additionally, to our knowledge, this is the first study %that addresses 
\BLUE{addressing}
market sensitivities focused on deep learning.
%%%%%%%%%%%%%%%%%%%%%%%%%%%%%%%%%%%%%%%%%%% 정리 %%%%%%%%%%%%%%%%%%%%%%%%%%%%%%%%%%%%%%%%%%%

% We would like to encourage you to list your keywords within
% the abstract section using the \keywords{...} command.
\end{abstract}
\section{Introduction}
\label{section:intro}

The assets under management (AUM) of index funds have increased steadily, accounting for a significant share (more than 30\%) of the AUM of U.S. equity funds in recent years~\cite{heath2022index}.
Index funds are a class of mutual fund whose returns closely resemble those of a predefined market index (e.g., the S\&P 500); the action of tracking such a market index is referred to as index tracking~\cite{oh2005using}.
Index-tracking strategies are instrumental to the management of index funds.

Depending on methodologies, index-tracking can be categorized into full replication and partial replication~\cite{kim2020index}.
Full replication involves constructing a portfolio with every constituent in the market index, each of which is market-capitalization-weighted.
Theoretically, full replication creates a complete portfolio with zero tracking error.
However, there exist numerous hurdles interrupting full replication, such as transaction costs, lack of trading volume of small-cap stocks, wide bid-ask spreads for less liquid assets, and investment constraints. 
\PURPLE{Thus, partial replication is utilized instead of full replication.}

Partial replication aims to deliver the market index's return, while the portfolio contains only a part of the market index constituents. 
% Nevertheless, 
It has a higher potential to reduce transaction costs because the portfolio holds comparatively fewer ``smaller and less liquid stocks.'' 
Furthermore, partial-replication methods are more flexible for building portfolios with strict constraints.
% \PURPLE{However, portfolio managers also face challenges in partial-replication management strategies due to fewer constituents.}
\PURPLE{However, portfolio managers also face challenges in partial-replication management strategies because containing fewer constituents might have a higher probability of the portfolio's returns deviating from the market index.}
% The fewer the constituents in the tracking portfolio, the more challenges portfolio managers experience.
Therefore, it is vital to develop practicable approaches to replicating a market index with comparatively fewer stocks to reduce tracking errors.

In developing a strategy to replicate a market index, considering portfolio constituents' market sensitivities to the market index is paramount \cite{oh2005using,chang2004evaluating,keim1999analysis}.
A financial instrument's market sensitivity to a market index indicates how sensitively the financial instrument reacts to the market index when the market index fluctuates \cite{sharpe1972risk,sharpe1995risk}. 
Market sensitivity is also called the $\beta$ of a financial instrument.
Generally, if a portfolio contains many high-$\beta$ financial instruments, its return may increase (decrease) more than that of the market index increases (decreases).
Therefore, the portfolio constituents' market sensitivities affect how closely the produced portfolio follows a market index, making it crucial to manage the portfolios' market sensitivities while building them.

However, impediments in estimating $\beta$ make it difficult for portfolio managers to control the $\beta$ of funds because $\beta$ is not directly observable but should be estimated or predicted based on historical data. 
There have been several approaches \cite{blume1975timevarying,ferson1993risk,pagan1980kalman,das2010market,engle2002dcc,engle2016garch} to handle this problem, but they have limitations: estimation of the limited number of stocks \cite{siegel1995option,buss2012option} and lack of out-of-sample tests \cite{hollstein2016beta,faff2000ts}.

To address these issues, we propose a deep-learning (DL) approach to predict market sensitivities utilizing historical price return data, and we substantiate \BLUE{our approach} utilizing the sliding window method \cite{hota2017time} that provides pseudo out-of-sample tests. Next, we suggest a mixed-integer linear programming (MILP) model that constructs index-tracking portfolios utilizing the predicted market sensitivities to handle the problem of partial replication. 
These methods' efficacy for prediction and portfolio construction is substantiated by our experiments on the Korean Stock Price Index 200 (KOSPI 200).

This study contributes as follows:
\vspace{-\topsep}
\begin{itemize}
\item[$\bullet$] To our knowledge, this is the first study to harness deep-learning models to predict market sensitivities.
\item[$\bullet$] We propose a method to construct partial-replication portfolios for index funds utilizing the predicted market sensitivities.
\item[$\bullet$] We present cumulative density function (CDF) transformation and a novel method to create target data, which are instrumental in training the prediction models. 
\item[$\bullet$] Using KOSPI 200, we corroborate our methods' efficacy in prediction and portfolio construction.
\item[$\bullet$] Our prediction models and portfolio construction methods, respectively, reduced historical estimations' prediction errors to about $57\%$ and the required number of stocks by half.
\end{itemize}

The remainder of this paper is organized as follows. First, we review the related work and background knowledge in sections 2 and 3, respectively. Then, section 4 elucidates our methods, and section 5 substantiates the proposed methods with experiments. Finally, we conclude the paper in section 6.

\section{Related Work}
\label{section:relwork}
%%%%%%%%%%%%%%%%%%%%%%%%%%%%%원문%%%%%%%%%%%%%%%%%%%%%%%%%%%%%
% Because the index tracking and prediction of the market sensitivity are the gist of our study, the prior literature was reviewed from the perspective of them.
%%%%%%%%%%%%%%%%%%%%%%%%%%%%%원문%%%%%%%%%%%%%%%%%%%%%%%%%%%%%
Since our study focuses on index tracking and prediction of market sensitivity, the literature review was conducted from this perspective.

\subsection{Index Tracking via Machine Learning}
%%%%%%%%%%%%%%%%%%%%%%%%%%%%%원문%%%%%%%%%%%%%%%%%%%%%%%%%%%%%
% In recent years, machine learning-based methods have been one of the most outstanding ways in partial replication for index tracking. 
% In \cite{oh2005using}, the authors proposed to exploit a genetic algorithm (GA) for index tracking. 
% % -> 베타 추정이 어떻게 이루어졌는지; 단순히 과거 데이터로 추정했는지 > 만약 그렇다면 한계점이 존재 -> 우리는 베타 추정 한다.
% The algorithm first uses some fundamental variables, such as standard error of portfolio beta, average trading amount, and average market capitalization, to select constituents of tracking portfolio. 
% Then, the portfolio weights of selected stocks are optimized through the GA.
% A heuristic method based on Hopfield Neural Network~\cite{hopfield1984neurons} is presented by~\cite{fernandez2007portfolio} in order to solve the general Markowitz mean-variance portfolio selection model with cardinality and bounding constraints. 
%%%%%%%%%%%%%%%%%%%%%%%%%%%%%원문%%%%%%%%%%%%%%%%%%%%%%%%%%%%%
Recently, machine-learning-based methods have become outstanding ways to conduct partial replication for index tracking. 
In \cite{oh2005using}, the authors proposed to exploit a genetic algorithm (GA) for index tracking. 
The algorithm first utilizes some fundamental variables to select tracking portfolio constituents. 
Then, through the GA, the algorithm optimized the portfolio weights of the selected stocks.
A heuristic approach utilizing Hopfield neural networks~\cite{hopfield1984neurons} was presented by~\cite{fernandez2007portfolio} to solve the generalized Markowitz mean-variance optimization with cardinality and bounding restrictions. 

Particularly, the adoption of deep neural networks (DNN) has broken new ground for index tracking.
\cite{kwak2021neural} introduced portfolio weights based on the output of a neural network that takes a fixed noise as input. However, portfolio weight determined by a fixed noise is hard to have a relationship with the stocks and index and cannot reflect the market situation.

Given time series data of index constituents, \cite{ouyang2019index} used a deep autoencoder to reconstruct the input and select stocks with minimum reconstruction losses as the index-tracking portfolio constituents.
Similarly, \cite{kim2020index} used a deep autoencoder to reconstruct the returns of index constituents. 
Stocks with the largest correlation coefficients or mutual information with the latent variable were selected as index-tracking portfolio constituents. 
The index-tracking portfolio's weight was calculated using a correlation coefficient.
While both abovementioned works utilized the deep autoencoder to connect individual stock information to market value, the interpretation of market information included by the latent variable has an obscure theoretical foundation.

In \cite{bradrania2022state}, the authors used a DNN to dynamically \BLUE{determine} the asset selection method (criterion) for the index-tracking portfolio conditioning on the market state and applied the cointegration method for building the index-tracking portfolio. 
\cite{zhengchen2020} proposed an index-tracking method for solving a partial-replication optimization problem with stochastic neural networks by employing reparametrization. It has limited reasoning capability for index tracking as well as a lack of comparison with other related methods.
% \cite{zhengchen2020} presented an index tracking approach with cardinality constraints to solve a partial replication optimization problem. They proposed a reparametrization method that enables it to be solved by simple first-order optimization methods, and then solve it with stochastic neural networks. Because they mainly focused on building a general approach to solve NP-hard optimization problem approximately, it has limited reasoning capability for index tracking as well as a lack of comparison with other related methods. 
% }

% %%%%%%%%%%%%%%%%%%%%%%%%%%%%%원문%%%%%%%%%%%%%%%%%%%%%%%%%%%%%
% To the best of our knowledge, no previous works have delved into the direct application and predictions of the market sensitivities in index-tracking portfolio constructions although the market sensitivities are instrumental to index tracking \cite{oh2005using}. In this study, we develop the deep learning models to predict the market sensitivities and utilize them directly by setting the net beta of the portfolio to be equal to the beta of the index.   
% % \ozumer{To the best of our knowledge, none of the previous works has focused on the methods that combines single factor models~~ blah blah}

% % -- 추가 https://www.tandfonline.com/doi/full/10.1080/0013791X.2022.2047851?casa_token=9-6_fpEJNK8AAAAA%3AdfviLCPhu8jn4EmOukpp6-EwUC9BW5IRC5ClBfKIz6Cihg5j0ehuK-GnZuvkL02KCZg-3VRm8zYAsg
% %%%%%%%%%%%%%%%%%%%%%%%%%%%%%원문%%%%%%%%%%%%%%%%%%%%%%%%%%%%%
To our knowledge, no previous works considered both the application and prediction of market sensitivities in index-tracking portfolio construction, although market sensitivities are instrumental to index tracking \cite{oh2005using}. This study develops DL models to predict market sensitivities and utilize them directly by setting the net $\beta$ of the portfolio to be equal to the $\beta$ of the index.   

\subsection{Estimation of Market Sensitivities} 

\label{sub:related_betas}

A simple approach to estimate $\beta$ involves generating the slope coefficient from a linear regression by utilizing historical time series of return data, which is referred to as a \textbf{historical estimation} in this paper. However, much evidence has shown that $\beta$ has time-varying properties \cite{blume1975timevarying,ferson1993risk}, which led to many other approaches based on generalized autoregressive conditional heteroskedasticity \cite{engle2002dcc,engle2016garch}, Kalman-filter \cite{pagan1980kalman,das2010market}, etc. According to \cite{hollstein2016beta,faff2000ts}, many of these intricate models attempting to capture the time-variation of $\beta$ perform better in training data but do not provide evidence from test data. Recently, a machine-learning-based approach~\cite{wolfgang2022ml} has been studied that estimates $\beta$ based on three different forecast model families: linear regressions, tree-based models, and neural networks. Because these studies focused on % the comparison of 
\BLUE{comparing} machine-learning methods, no DL method other than multi-layer perceptrons (MLP) has been reviewed. % This study focuses on DL, utilizes other DL architectures, and examines their practical efficacy in terms of index tracking using the sliding window method \cite{hota2017time} to evaluate properly.

Option implied estimation approaches have also been widely studied~\cite{siegel1995option,buss2012option}. These approaches have a 
% big 
\BLUE{significant} advantage as they can consider forward-looking information from options markets, but only limited stocks that have option derivatives can be analyzed. 
% This limitation could be critical to achieving the main purpose of this study because our final goal is not to estimate $\beta$ itself but create a portfolio that tracks the underlying index well using the estimated $\beta$. 
\ORANGE{This limitation could be fatal to achieving the main goal of this study because index tracking requires a greater number of market sensitivities.} 
Moreover, \cite{hollstein2016beta} states that fully option implied approaches~\cite{chang2011option,sr2005option,kks2014option} have substantial errors that they cannot produce negative values.  

\ORANGE{To address these limitations, this study proposes a DL approach for time-varying $\beta$ with canonical DL architectures and examines efficacy in terms of index tracking using the sliding window method \cite{hota2017time} for adequate evaluation.}

\section{Preliminaries}
\label{section:prelim}

This section explicates an overview of a single-factor model and introduces the basics of several DL models. 

\subsection{Notations}
\PURPLE{Let $t\in\mathbb{Z}$ be a time step to index each trade date in sequence. We assume that an investor can trade a financial instrument $i$ in an investment universe set $S_t$ at $t$ and that every trade for $i$ at $t$ is executed at close price $p_{i,t}$. 
% Next, we denote $p_{i,t}\in \mathbb{R}$ as the close price of $i$ at $t$. 
From these, we define $r_{i,t:t'}\in \mathbb{R}$ as the return of $i$ from $t$ to $t'$ as (\ref{eqn:def_ret}). Additionally, we suppose a market index $m$ is given, whose return is defined as $r_{m,t:t'}$ like $r_{i,t:t'}$.
\begin{align}
\label{eqn:def_ret}
r_{i,t:t'}= p_{i,t'}/p_{i,t}-1    
\end{align}}
\subsection{Single-Factor Model}
Let $f_{t:t'}\in \mathbb{R}$ be a factor that affects $r_{i,t:t'}$. Then, pursuant to the single-factor model, we assume that $r_{i,t:t'}$ can be expressed by $f_{t:t'}$ as (\ref{eqn:factor_model}) \cite{luenberger2009investment}. \(\beta_i,\alpha_i,\epsilon_i \in \mathbb{R}\) are the market sensitivity (or slope), excess return (or intercept), and error of the factor  $f_{t:t'}$ for \(i\), respectively. 
\begin{align}\label{eqn:factor_model}
r_{i,t:t'} = \alpha_i + \beta_i f_{i,t:t'}  + \epsilon_i
\end{align}
In this study, we consider only the return of a given index \(r_{m,t:t'}\) as a factor $f_{t:t'}$ and assume that \(\beta_i,\alpha_i,\epsilon_i\) vary as time \(t\) passes, as in (\ref{eqn:factor_model2}):
\begin{align}\label{eqn:factor_model2}
r_{i,t:t'} = \alpha_{i,t:t'} + \beta_{i,t:t'} r_{m,t:t'} + \epsilon_{i,t:t'}
\end{align}\PURPLE{where $\alpha_{i,t:t'}, \beta_{i,t:t'}, \epsilon_{i,t:t'} \in \mathbb{R}$}.
Then, the return of a portfolio \(\pi\) whose financial instruments are each initially weighted by \(w_{i,t}\in\mathbb{R}\) at $t$ and not sold nor bought until $t'$ becomes (\ref{eqn:factor_model3_1}):

%\begin{align}
%\label{eqn:factor_model3_0}
%r_{\pi, t:t'} &=\sum_{i \in S_t} w_{i, t} r_{i, t:t'} \\
%\label{eqn:factor_model3_1}
%&= \sum_{i\in S_t} (w_{i, t} \alpha_{i, t:t'} \\
%\label{eqn:factor_model3_2}
%& + w_{i, t} \beta_{i, t:t'} r_{m, t:t'} \\
%\label{eqn:factor_model3_3}
%& + w_{i, t} \epsilon_t).
%\end{align}

\begin{align}
\label{eqn:factor_model3_0}
r_{\pi, t:t'} &=\sum_{i \in S_t} w_{i, t} r_{i, t:t'} \\
\label{eqn:factor_model3_1}
&= \sum_{i\in S_t} (w_{i, t} \alpha_{i, t:t'} + w_{i, t} \beta_{i, t:t'} r_{m, t:t'} + w_{i, t} \epsilon_t)
\end{align}

\subsection{Deep-Learning Models}

DL is a method for discovering the connection between multiple features and the knowledge behind the connection, according to \cite{zhang2018definition}. To extract the relationship and knowledge, several well-known DL models were suggested, and we utilize four of them: MLP, long-short term memory (LSTM) \cite{hochreiter1997long}, gated-recurrent unit (GRU) \cite{cho2014learning}, and Transformer \cite{vaswani2017attention}.

MLP consists of several connected layers of artificial neurons that take inputs, multiply weights to them, sum them up, and output the sum after it passes a nonlinear activation \cite{gardner1998artificial}. Because it has nonlinear activation functions, it can extract nonlinear features from the data. Then, to handle sequential data, LSTM and GRU were suggested. In LSTM, the cells utilize long-term information as well as short-term or new inputs. Similarly, GRU utilizes long-term memory but has a simpler design, so it has fewer learnable parameters \cite{fu2016using}. In place of sequence-to-sequence structures like LSTM and GRU, Transformer utilizes a multi-head attention mechanism to extract features. This study employs the encoder layers of the Transformer, but for simplicity, we refer to them as Transformer or Trans.

% \section{Problem Formulation}
% \label{section:problem}
% \input{text/problem_formulation}

% \begin{figure*}[htp]
%   \centering
%   \includegraphics[width=\linewidth]{img/Timeline.PDF}
%   \caption{Timeline of Portfolio Updates}
%   % \Description{}
%   \label{fig:Update Timeline}
% \end{figure*}

\begin{figure*}[pt]
  \centering
  \includegraphics[width=0.95\linewidth]{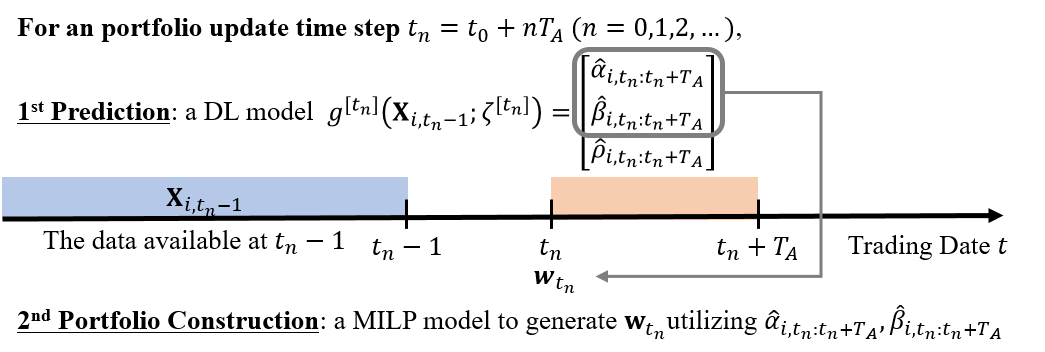}
  \caption{Timeline of Portfolio Updates }
  % \Description{}
  \label{fig:Update Timeline}
\end{figure*}
\section{Proposed Methods}
\label{section:method}

\subsection{Overview of Our Methods}
\label{subsec:overview_methods}
 \PURPLE{To partially replicate a market index $m$,} 
 our proposed methods solve two problems: (\RN{1}) \textbf{prediction} of \PURPLE{\(\alpha_{i,t_n:t_n+T_A},\beta_{i,t_n:t_n+T_A} \in\mathbb{R}\)} of the single-factor models, and (\RN{2}) \textbf{portfolio} \textbf{construction} determining  \PURPLE{$\textbf{w}_{t_n}=[w_{i,t_n}]_{i\in S_{t_n}} \in [0,1]^{\vert S_{t_n} \vert} $}.
The DL models are harnessed to the former, and MILP is employed for the latter (Fig.~\ref{fig:Update Timeline}).
Specifically, \PURPLE{at each $t_n=t_0+nT_A$ ($n=0,1,2,...$ $; T_A\in\mathbb{N}$ ), $g^{[t_n]}$ is newly trained and predicts $\{ (\hat{\alpha}_{i,t_n:t_n+T_A},\hat{\beta}_{i,t_n:t_n+T_A}): $$~i $$\in$$ S_{t_n}\}$; our MILP model takes the predicted values as parameters} 
\PURPLE{and outputs $\textbf{w}_{t_n}$; and our portfolio is updated to $\textbf{w}_{t_n}$.
$T_A$ is a given trading period (in the experiment, \(T_A=21\) trading days, assumed to be a month).}
\BLUE{Note that the data at $t_n$ must not be used for training $g^{[t_n]}$ and generating $\textbf{w}_{t_n}$ to avoid a look-ahead bias\footnote{
\BLUE{
Look-ahead bias is a bias caused by utilizing data that are unavailable when constructing portfolios \cite{zhou2014active,isichenko2021quantitative}. 
% For example, if an investing strategy utilizes the a-month-later stock prices for constructing tomorrow's portfolios, then it has a look-ahead bias.
For example, if an investing strategy uses data generated at time $t$ to build a portfolio at $t$, then it has a look-ahead bias because the data generated at time $t$ cannot be delivered to an investor at $t$ due to delivery time.
}
} \cite{zhou2014active,isichenko2021quantitative}. 
The data up to $t_n-1$ are only available for training $g^{[t_n]}$ and generating $\textbf{w}_{t_n}$. 
%That is, the data at $t_n$ are unavailable for training $g^{[t_n]}$ and generating $\textbf{w}_{t_n}$.
}
% it is assumed that, a portfolio is first constructed at \(t_0\) and updated periodically with a given trading period \(T_A\in\mathbb{N}\) (in the experiment, \(T_A=21\) trading days, assumed to be a month) at close prices. \PURPLE{At each $t_n=t_0+nT_A$ ($n=0,1,2,...$ ), $g^{[t_n]}$ is newly trained and predicts $\{ (\hat{\alpha}_{i,t_n:t_n+T_A},\hat{\beta}_{i,t_n:t_n+T_A}): $$~i $$\in$$ S_{t_n}\}$
% the portfolio is updated to $\textbf{w}_{t_n}$ our MILP model determines. 
% To do so, $g^{[t_n]}$ is newly trained at every $t_n$ and predicts $\{ (\hat{\alpha}_{i,t_n:t_n+T_A},\hat{\beta}_{i,t_n:t_n+T_A}): $$~i $$\in$$ S_{t_n}\}$ that our MILP model takes as inputs and outputs $\textbf{w}_{t_n}$
% % our MILP model takes as inputs $\{ (\hat{\alpha}_{i,t_n:t_n+T_A},\hat{\beta}_{i,t_n:t_n+T_A}): $$~i $$\in$$ S_{t_n}\}$ that our prediction model $g^[t_n]$ predicts,

\subsubsection{Prediction of Market Sensitivities}
\label{subsec:ps}

\PURPLE{
For a portfolio weight $\textbf{w}_{t_n}$, our prediction model $g^{[t_n]}$ parameterized by $\zeta^{[t_n]}$ predicts $\hat{\alpha}_{i,t_n:t_n+T_A},$ $\hat{\beta}_{i,t_n:t_n+T_A}$  with an input $\textbf{X}_{i,t_n-1}$, as shown in Fig.~\ref{fig:Update Timeline}. Without loss of generality, we define $\textbf{X}_{i,t-1}$ as the data available at $t-1$. Additionally, $g^{[t_n]}$ outputs a residual \(\hat{\rho}_{i,t_n:t_n+T_A}\) in replacement of error $\epsilon_{i,t_n:t_n+T_A}$. We intend $g^{[t_n]}$ to learn better representations by providing the values of \(\hat{\rho}_{i,t_n:t_n+T_A}\) when training it.}
% To do so, in Fig.~\ref{fig:Update Timeline}, the data in time period \([t-T_B,t-1]\) denoted by \(\textbf{X}_{it}\) are utilized for the DL models to predict \(\beta_{it},\alpha_{it} \in\mathbb{R}\) for \(r_{it}, r_{mt}\). $T_B$ determines how long data will be utilized. \(r_{it}, r_{mt}\) are the returns of an instrument \(i\) and index \(m\), respectively, from their close prices at \(t\) to \(t+T_A\).
Then, the objective of the prediction is to minimize the prediction errors (PE) of one-factor models, which are formulated as follows. 

\begin{align}
\label{eqn:obj_pred_1}
\min_{\zeta^{[t_n]}} &~PE  =\mathbb{E}[ ~ (r_{i,t_n:t_n+T_A} - \hat{r}_{i,t_n:t_n+T_A})^2~ ] \\
\label{eqn:obj_pred_2}
s.t. &~\hat{r}_{i,t_n:t_n+T_A} = \hat{\beta}_{i,t_n:t_n+T_A} r_{m,t_n:t_n+T_A} +\hat{\alpha}_{i,t_n:t_n+T_A} + \hat{\rho}_{i,t_n:t_n+T_A}\\
\label{eqn:obj_pred_3}
&~ [\hat{\alpha}_{i,t_n:t_n+T_A},\hat{\beta}_{i,t_n:t_n+T_A},\hat{\rho}_{i,t_n:t_n+T_A}]^T = g^{[t_n]}(\textbf{X}_{i,t_n-1};\zeta^{[t_n]})
\end{align}

\subsubsection{Construction of Partially Replicated Portfolio}
\label{subsec:cprp}

\PURPLE{
To output portfolio weight $\textbf{w}_{t_n}$ partially replicating a market index $m$, our portfolio construction model minimizes tracking error (TE) as (\ref{eqn:obj_port_}), where \(\vert\vert{\tiny\bullet} \vert\vert\) is a norm.} 
% To construct the constituent weights of a portfolio,  \(\boldsymbol{w}_t=[w_{it}] \in [0,1]^{|S|}\), we minimize tracking errors (TE) as Equation~(\ref{eqn:obj_port_}), where \(||{\tiny\bullet} ||\) is a norm.
Although there are many definitions of TE \cite{pope1994discovering,roll1992mean}, the one used in this study is (\ref{eqn:obj_port_}). \PURPLE{To achieve (\ref{eqn:obj_port_}), our MILP model takes predicted $\hat{\alpha}_{i,t_n:t_n+T_A},$ $\hat{\beta}_{i,t_n:t_n+T_A}$ as inputs and utilizes them as its parameters, which will be elucidated in subsection \ref{subsec:port_constr}.} 

Additionally, we want our portfolio to contain fewer financial instruments than the market index \(m\) in number. 
Let \(N^* \in \mathbb{N}\) be the limit on the number of financial instruments in the portfolio.
We also set \({u}_{i,t_n} \in \{0,1\}\) as an indicator variable showing if the portfolio at $t$ includes a financial instrument \(i\) \PURPLE{(i.e., if $w_{i,t_n}\neq0$, $u_{i,t_n}$$=$$1$)}.
Then, a constraint (\ref{eqn:const_num}) restricts the number of financial instruments in the portfolio \cite{canakgoz2009mixed}.
% \begin{align}
% \label{eqn:obj_port_}
% % &minimize\quad &{TE} (r_{\pi,t:t+T_A}, r_{m,t:t+T_A})
% &minimize\quad &{TE} =\mathbb{E}[~ ||r_{\pi t}-r_{mt}|| ~] \\
% \label{eqn:const_num}
% &where\quad &\sum_{i\in S_t} u_{it} \leq N^*
% \end{align}
\PURPLE{
\begin{align}
\label{eqn:obj_port_}
% &minimize\quad &{TE} (r_{\pi,t:t+T_A}, r_{m,t:t+T_A})
& \min \quad &{TE} =\mathbb{E}[ \vert\vert r_{\pi,t_n:t_n+T_A}-r_{m,t_n:t_n+T_A} \vert\vert ] \\
\label{eqn:const_num}
& s.t.\quad &\sum_{i\in S_{t_n}} u_{i,t_n} \leq N^*
\end{align}
}
\subsubsection{Construction of Partially Replicated Portfolio}
\label{subsec:estimation_metrics}
\PURPLE{
We use the simple means  as $\hat{PE}$ and $\hat{TE}$ in (\ref{eqn:PE_est}) and (\ref{eqn:TE_est}), respectively, to estimate the expectation $\mathbb{E}$ in (\ref{eqn:obj_pred_1}) and (\ref{eqn:obj_port_}) when training or evaluating our models for the two tasks. $V=\{(i,t) : t \in \mathfrak{T} ~~and~~i \in {S_t}'\}$ is a batch given sets of some time steps $\mathfrak{T}$ and some financial instruments ${S_t}'$.
\begin{align}
\label{eqn:PE_est}
&\hat{PE}={1 \over \vert V \vert} \sum_{(i,t)\in V} (r_{i,t:t+T_A} - \hat{r}_{i,t:t+T_A})^2
% &\hat{PE}={1 \over \vert V \vert} \sum_{(i,t)\in V} (r_{i,t:t+T_A} - \hat{\beta}_{i,t:t+T_A} r_{m,t:t+T_A} -\hat{\alpha}_{i,t:t+T_A})^2
\end{align}\begin{align}
% \label{eqn:PE_est}
% &\hat{PE}={1 \over |V|} \sum_{v\in V} (r_{it} - \hat{\beta}_{it} r_{mt} -\hat{\alpha}_{it})^2 \\
\label{eqn:TE_est}
&\hat{TE} = {1 \over {\vert \mathfrak{T}\vert }} \sum_{t\in \mathfrak{T}} (r_{\pi,t:t+T_A}-r_{m,t:t+T_A})^2
\end{align}
}

\begin{figure}[pt]
  \centering
  \includegraphics[width=0.95\linewidth]{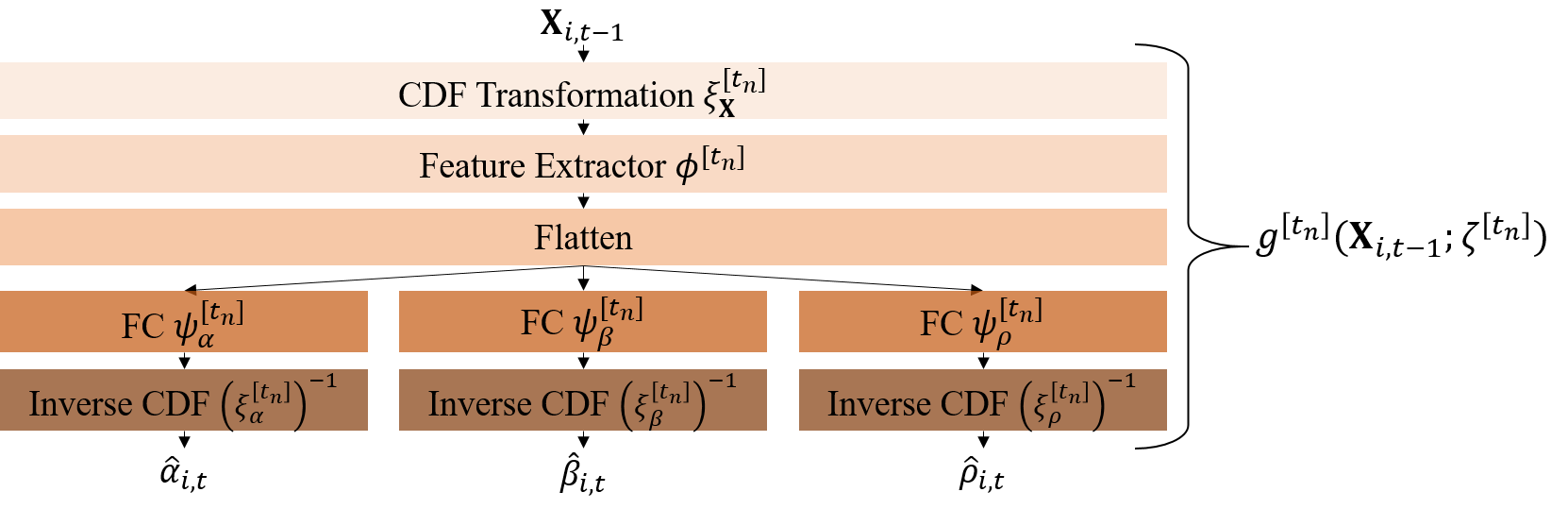}
  \caption{Architecture of Deep-Learning Models $g^{[t_n]}(\textbf{X}_{i,t-1};\zeta^{[t_n]})$}
  % \Description{...}
  \label{fig:dl_arc}
\end{figure}
\subsection{Prediction of Market Sensitivities via Deep Learning Models}
\label{subsec:deepmodel}

To predict \(\hat{\alpha}_{i,t_n:t_n+T_A},\hat{\beta}_{i,t_n:t_n+T_A}\) of the single-factor models used in the portfolio construction step, we train the DL model $g^{[t_n]}$ in a supervised manner. The overall architecture of our models is depicted in Fig.~\ref{fig:dl_arc}, which is CDF transformation ($\xi_{\textbf{X}}^{[t_n]}$), feature extractor ($\phi^{[t_n]}$), fully-connected (FC) layers (\(\psi_{\tiny\bullet}^{[t_n]}\)), and inverse CDF transformations ($(\xi^{[t_n]}_{\tiny\bullet})^{-1}$).
Note that subscripts of the four components do not have $i$, which means that a single model is trained and used to predict all financial instruments at $t_n$.

\subsubsection{Architecture of Deep Learning Models}
\label{subsec:arc_dl}

First, when input \(\textbf{X}_{i,t-1}\) passes into the neural network, its distribution is reshaped by the CDF transformation to expedite the learning process of the models. 
Details of the transformation are explicated in the next subsection. 
Second, the feature extractor receives the transformed input data. 
The feature extractor's primary function is finding appropriate representations for prediction. So, any kind of (sub-)differentiable architecture of the neural networks can be harnessed to it. 
Inside the feature extractors, dropout layers are interposed to attenuate over-fitting.
The retrieved representations are then flattened before entering the FC layers \PURPLE{\(\psi_{\tiny\bullet}\), whose last activation function is a sigmoid function. 
Finally, at the inverse CDF transformation, the outputs of  \(\psi_{\tiny\bullet}\) are turned into actual values of \(\hat{\beta}_{i,t:t+T_A}, \hat{\alpha}_{i,t:t+T_A}, \hat{\rho}_{i,t:t+T_A}\).}

\begin{figure}[pt]
  \centering
  \includegraphics[width=0.6\linewidth]{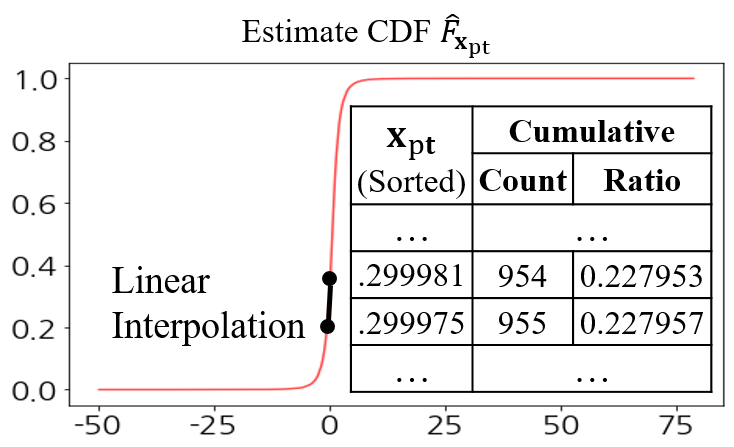}
  \caption{Cumulative Distribution Function (CDF) Transformation; Suppose input data \(X_t^{train}=[\textbf{x}_{1t} \textbf{x}_{2t}~...~\textbf{x}_{pt}~...~]\), where \(\textbf{x}_{pt}\) is the \(p\)-th feature vector whose elements are samples. 
% As the left histogram in Figure \ref{fig:cdf_trans}, the samples of \(\textbf{x}_{pt}\) are widely distributed, for example. 
We find the cumulative mass function (CMF) of the samples of \(\textbf{x}_{pt}\) by sorting and counting. 
Then, the CMF of \(\textbf{x}_{pt}\) is linearly interpolated to approximate the CDF of \(\textbf{x}_{pt}\). Denote the approximated CDF of \(\textbf{x}_{pt}\) as \(\hat{F}_{\textbf{x}_{pt}}({\tiny\bullet})\). 
% As shown at the right histogram, the distribution of \(\hat{F}_{\textbf{x}_{pt}}({\textbf{x}_{pt}})\) is uniformly distributed. 
Then, \(\hat{F}_{\textbf{x}_{pt}}({\textbf{x}_{pt}})\) becomes uniformly distributed. }
  % \Description{...}
  \label{fig:cdf_trans}
\end{figure}

\subsubsection{CDF Transformation}
\label{subsec:cdf}

To expedite the optimization of DL models, we apply CDF transformations to the input and output data.
CDF transformations transmute any continuous distributions on the real line into a uniform distribution $(0, 1)$~\cite{casella2021statistical}. 
The input and output data in our study have wide-ranged distribution, %which makes
\ORANGE{making} it challenging for the models to learn. 
The CDF transformation addresses this issue by scaling down the data to the uniform distribution. 
This CDF transformation is applied variable-wisely as Fig.~\ref{fig:cdf_trans}. 
Note that the approximated CDF's inverse exists because the approximated CDF is a strictly increasing function.

% To avoid the look-ahead bias\footnote{
% \PURPLE{Look-ahead bias is a bias caused by an investing strategy utilizing future information or data that are unavailable when constructing portfolios \cite{zhou2014active,isichenko2021quantitative}. For example, if an investing strategy utilizes the a-month-later stock prices for constructing tomorrow's portfolios, then it has a look-ahead bias.}
% } \cite{zhou2014active,isichenko2021quantitative}, we preclude the validation and test datasets when \ORANGE{defining $\xi_{\textbf{X}}^{[t_n]}, \xi^{[t_n]}_{\tiny\bullet}$} as Fig.~\ref{fig:cdf_trans}.  

\ORANGE{We respectively define $\xi_{\textbf{X}}^{[t_n]}, \xi^{[t_n]}_{\tiny\bullet}$} as Fig.~\ref{fig:cdf_trans} based on input and output values of only a train dataset, without validation and test datasets, to avoid a look-ahead bias \cite{zhou2014active,isichenko2021quantitative}. 
Next, the defined $\xi_{\textbf{X}}^{[t_n]}, \xi^{[t_n]}_{\tiny\bullet}$ transmute the whole train and validation datasets. 
% Then, we train the DL models on both CDF-transformed input and output space; and apply the inverse CDF transformation to outputs of the DL model.
Based on both CDF-transformed input and output data, we train the DL models \ORANGE{without $\xi_{\textbf{X}}^{[t_n]},~(\xi^{[t_n]}_{\tiny\bullet})^{-1}$}. \ORANGE{Finally, in the inference phase, the trained DL models use \ORANGE{$\xi_{\textbf{X}}^{[t_n]},~(\xi^{[t_n]}_{\tiny\bullet})^{-1}$} and take non-transformed inputs.}

\begin{figure}[pt]
  \centering
  \includegraphics[width=0.91\linewidth]{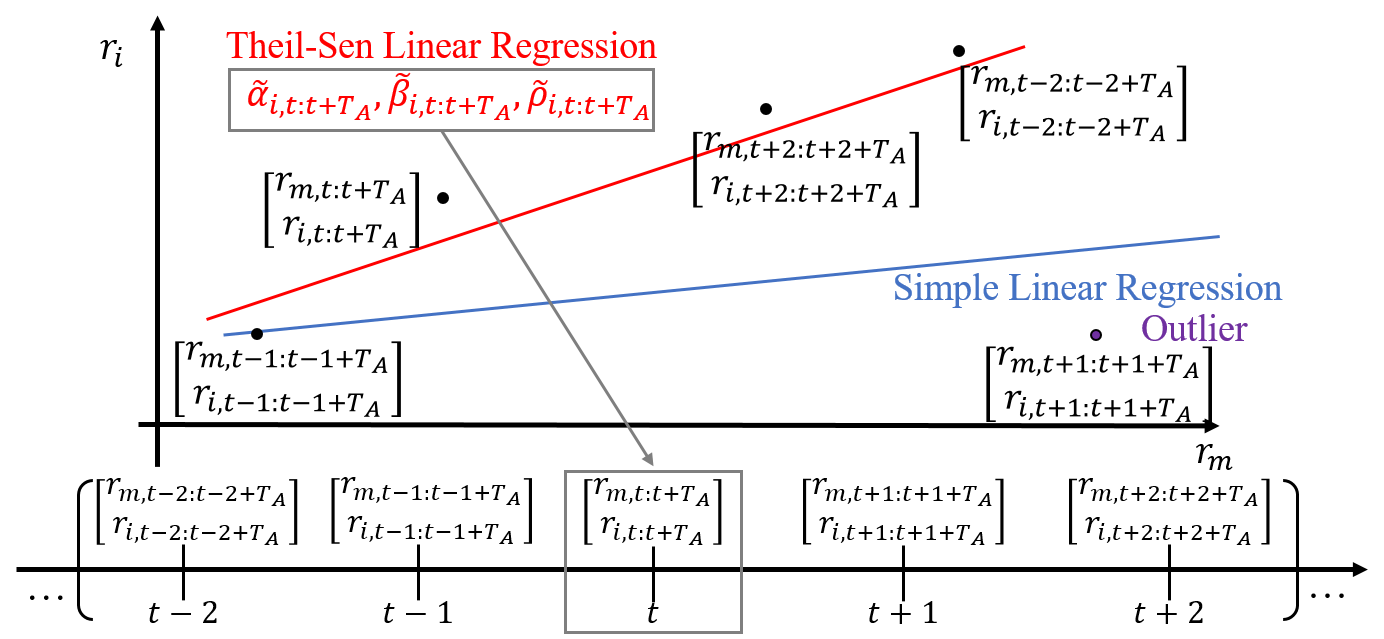}
  \caption{Theil-Sen Regression and Estimation of $\hat{\beta}_{i,t:t+T_A}, \hat{\alpha}_{i,t:t+T_A}, \hat{\rho}_{i,t:t+T_A}$. If $(r_{m,t+1:t+1+T_A},$ $r_{i,t+1:t+1+T_A})$ is an outlier in this figure, the simple linear regression would lean toward the outlier. However, if the Theil-Sen linear regression is applied, the regressed line leans less toward the outlier.}
  % \Description{...}
  \label{fig:theil_sen}
\end{figure}

\subsubsection{Supervised Learning}
\label{subsec:sl}

% \ORANGE{This study} 
We design a supervised-learning problem with the target data of \(\tilde{\beta}_{i,t:t+T_A}, \tilde{\alpha}_{i,t:t+T_A}, \tilde{\rho}_{i,t:t+T_A}\), generated by Theil-Sen linear regression \cite{wang2005asymptotics,dang2008theil} to train the aforementioned DL models, as in Fig.~\ref{fig:theil_sen}. 
\ORANGE{However}, if  $r_{i,t:t+T_A}$, $r_{m,t:t+T_A} $ of a single time-step \(t\) were utilized to estimate $\tilde{\beta}_{i,t:t+T_A}, \tilde{\alpha}_{i,t:t+T_A}, \tilde{\rho}_{i,t:t+T_A}$ with (\ref{eqn:factor_model2}), there are infinitely many pairs of them. This is because there are three variables, but one equation is provided. 

To address this, we utilize $r_{i,\tau:\tau+T_A}, r_{m,\tau:\tau+T_A}$ ($\tau \in \mathbb{Z},t-T_C \leq \tau \leq t+T_C$) to estimate \(\tilde{\beta}_{i,t:t+T_A}, \tilde{\alpha}_{i,t:t+T_A}, \tilde{\rho}_{i,t:t+T_A}\) with (\ref{eqn:factor_model2}); then, there are (\(2T_C+1\)) equations with three variables. 
Next, a linear regression method is applied to estimate \(\tilde{\beta}_{i,t:t+T_A}, \tilde{\alpha}_{i,t:t+T_A}, \tilde{\rho}_{i,t:t+T_A}\). 
\PURPLE{Note that $r_{i,t''-T_A:t''}$ is utilized to compute $\Tilde{\beta}_{i,t''-T_A-T_C:t''-T_C}$ for $t''=t+T_C+T_A$, and this property is utilized in \ref{subsec:data_splitting}.}

Because the stock market data have a lot of noise \cite{pafka2003noisy}, if a simple linear regression is applied, outliers can negatively impact the estimation of parameters. 
To alleviate the negative effects of outliers, we employ Theil-Sen linear regression, which is a robust linear regression method.

\subsection{Portfolio Construction}
\label{subsec:port_constr}
\subsubsection{Objective Function of Partial-Replication Portfolios}
\PURPLE{
First, we define notations to explain our partial-replication method.
Let $w_{i,t}^m$ be the weight of a financial instrument \(i\) in the market index \(m\) at $t$. 
% Because $w_{i,t_n}^m$ is unavailable when $w_{i,t_n}$ is constructed, we consider $w_{i,t-1}^m$ at $t-1$, not $t$, to prevent the look-ahead bias \cite{zhou2014active,isichenko2021quantitative}. 
We define market-weighted averages~$\overline{\alpha}_{m,t:t+T_A},~\overline{\beta}_{m,t:t+T_A},~\overline{\epsilon}_{m,t:t+T_A}$ as:~$\overline{\alpha}_{m,t:t+T_A}=\sum_{i \in S} w_{i,t}^m \hat{\alpha}_{i,t:t+T_A}$, $\overline{\beta}_{m,t:t+T_A}$ $=$ $\sum_{i \in S} w_{i,t}^m \hat{\beta}_{i,t:t+T_A}$, $\overline{\epsilon}_{m,t:t+T_A}$ $=$ $\sum_{i \in S} w_{i,t}^m \hat{\epsilon}_{i,t:t+T_A}$.
% \begin{align}
% \label{eqn:def_abar} \overline{\alpha}_{m,t:t+T_A}&=\sum_{i \in S} w_{i,t}^m \hat{\alpha}_{i,t:t+T_A}, \\
% \label{eqn:def_bbar} \overline{\beta}_{m,t:t+T_A}&=\sum_{i \in S} w_{i,t}^m \hat{\beta}_{i,t:t+T_A}, \\
% \label{eqn:def_ebar} \overline{\epsilon}_{m,t:t+T_A}&=\sum_{i \in S} w_{i,t}^m \hat{\epsilon}_{i,t:t+T_A}~.
% \end{align}
}

\PURPLE{
Now, utilizing the predicted $\hat{\alpha}_{i,t_n:t_n+T_A}, \hat{\beta}_{i,t_n:t_n+T_A}$, we decompose the objective of partial-replication portfolios in (\ref{eqn:obj_port_}). 
The difference of the returns in (\ref{eqn:obj_port_}) becomes  (\ref{eqn:tracking_error2}) by (\ref{eqn:factor_model3_0}) and can be rewritten into (\ref{eqn:tracking_error3})--(\ref{eqn:tracking_error5}) by (\ref{eqn:factor_model3_1}) and the definitions of $\overline{\alpha}_{m,t:t+T_A}, \overline{\beta}_{m,t:t+T_A}, \overline{\epsilon}_{m,t:t+T_A}$.
\begin{align}\label{eqn:tracking_error2}
r_{\pi,t_n:t_n+T_A}-r_{m,t_n:t_n+T_A} 
&=\sum_{i \in S_{t_n}} w_{i,t_n} r_{i,t_n:t_n+T_A} - \sum_{i \in S_{t_n}} w_{i,t_n}^m  r_{i,t_n:t_n+T_A} \\
\label{eqn:tracking_error3}
&=(\sum_{i\in S_{t_n}} w_{i,t_n} \hat{\beta}_{i,t_n:t_n+T_A} -  \overline{\beta}_{m,t_n:t_n+T_A}) r_{m,t_n:t_n+T_A} \\
\label{eqn:tracking_error4}
&~~~+ \sum_{i\in S_{t_n}} w_{i,t_n} \hat{\alpha}_{i,t_n:t_n+T_A}-  \overline{\alpha}_{m,t_n:t_n+T_A} \\
\label{eqn:tracking_error5}
&~~~+ \sum_{i\in S_{t_n}} w_{i,t_n} \hat{\epsilon}_{i,t_n:t_n+T_A}-  \overline{\epsilon}_{m,t_n:t_n+T_A}
\end{align}
} 

\PURPLE{
If we construct a portfolio with the weights $w_{i,t_n}$ satisfying constraints (\ref{eqn:const_bbar}) and (\ref{eqn:const_abar}), we do not need to predict the market index's return $r_{m,t:t+T_A}$ when replicating $r_{m,t:t+T_A}$. 
This is because the coefficient of $r_{m,t_n:t_n+T_A}$ in (\ref{eqn:tracking_error3}) becomes zero if (\ref{eqn:const_bbar}) is satisfied. 
That is, the market index's return $r_m$ does not affect our partial-replication objective (\ref{eqn:obj_port_}). 
Hence, those constraints allows us not to predict the market index's return $r_{m,t:t+T_A}$ when we are replicating it. Moreover, (\ref{eqn:tracking_error4}) becomes zero if (\ref{eqn:const_abar}) is satisfied. Hence, our objective function (\ref{eqn:obj_port_}) with constraints (\ref{eqn:const_bbar}) and (\ref{eqn:const_abar}) becomes \(\mathbb{E}[\vert\vert\sum_{i\in S} w_{i,t_n} \hat{\epsilon}_{i,t_n:t_n+T_A}-  \overline{\epsilon}_{m,t_n:t_n+T_A}\vert\vert]\). 
\begin{align}
\label{eqn:const_bbar} \sum_{i\in S_{t_n}} w_{i,t_n} \hat{\beta}_{i,t_n:t_n+T_A} = \overline{\beta}_{m,t_n:t_n+T_A}  \\
\label{eqn:const_abar} \sum_{i\in S_{t_n}} w_{i,t_n} \hat{\alpha}_{i,t_n:t_n+T_A} = \overline{\alpha}_{m,t_n:t_n+T_A}  
\end{align}
}

However, $\overline{\alpha}_{m,t_n:t_n+T_A}$ and $\overline{\beta}_{m,t_n:t_n+T_A}$ are unavailable at the time when we are determining $w_{i,t_n}$. This is because $w_{i,t_n}$ is determined before $t_n$, but $w^m_{i,t_n}$ is available after $t_n$ and $\overline{\alpha}_{m,t_n:t_n+T_A}$, $\overline{\beta}_{m,t_n:t_n+T_A}$ use $w^m_{i,t_n}$. To address this, we additionally assume as follows: 
$\overline{\alpha}_{m,t_n:t_n+T_A} \approx \sum_{i \in S} w_{i,t_n-1}^m \hat{\alpha}_{i,t_n:t_n+T_A},$ 
$\overline{\beta}_{m,t_n:t_n+T_A} \approx \sum_{i \in S} w_{i,t_n-1}^m \hat{\beta}_{i,t_n:t_n+T_A}.$
% $$\overline{\epsilon}_{m,t_n:t_n+T_A} &\approx \sum_{i \in S} w_{i,t_n-1}^m \hat{\epsilon}_{i,t_n:t_n+T_A}~.$$
Then, we can construct a portfolio that is weighted by $w_{i,t_n-1}$ and that satisfies (\ref{eqn:const_bbar}) and (\ref{eqn:const_abar}) at $t_n$.

\PURPLE{
Now, our portfolio aims to minimize $\mathbb{E}[ \vert\vert\sum_{i\in S} w_{i,t_n} \hat{\epsilon}_{i,t_n:t_n+T_A}-  \overline{\epsilon}_{m,t_n:t_n+T_A}\vert\vert ]$ with constraints (\ref{eqn:const_bbar}) and (\ref{eqn:const_abar}). }
However, it is difficult to predict errors accurately since they can be a combination of the influence of other factors with random noise. 
Hence, in lieu of minimizing \(\mathbb{E}[~ \vert\vert \sum_{i\in S_{t_n}} w_{i,t_n} \epsilon_{i,t_n:t_n+T_A}-  \overline{\epsilon}_{m,t_n:t_n+T_A}\vert\vert ~]\), we make $\textbf{w}_{t_n}$ close to $\textbf{w}_{t_n-1}^m$ at each $t_n$ as (\ref{eqn:surrogate_obj0}), 
% Here, \(\textbf{w}_t, \textbf{w}_{t-1}^m\in [0,1]^{\vert S \vert}\) are vectors, whose components are \(w_{it},w_{it}^m\), respectively. 
because if \(\textbf{w}_{t_n}\) and $\textbf{w}_{t_n-1}^m$ become similar, the portfolio will share similar unexplained factors or errors. \PURPLE{Additionally, the difference in (\ref{eqn:surrogate_obj0}) is a broadcasting difference. That is, if $i^*$ is not in $S_{n,t_n-1}$ or $S_{n,t_n}$, then $w_{i^*,t_n-1}^m$ or $w_{i^*,t_n}$ is defined as zero, respectively, and the difference is calculated stock-wisely (i.e., $\textbf{w}_{t_n}-\textbf{w}_{t_n-1}^m$ $\in$ 
$\mathbb{R}^{\vert S_{t_n} \cup S_{t_n-1}\vert}$).}
% . Hence, the dimension of $\textbf{w}_{t_n}-\textbf{w}_{t_n-1}^m$ is $\vert S_{t_n} \cup S_{t_n-1}\vert$.
\begin{align} 
\label{eqn:surrogate_obj0}
minimize ~~~  \vert\vert\textbf{w}_{t_n}-\textbf{w}_{t_n-1}^m\vert\vert
\end{align} 
When applying the norm to the objective function \(||\textbf{w}_{t_n}-\textbf{w}_{t_n-1}^m||\), we consider the sum of the 1-norm and infinity-norm, (\ref{eqn:surrogate_obj}), \PURPLE{where $S=S_{t_n} \cup S_{t_n+1}$.} 
\PURPLE{The reason why the two norms are selected is that they can be linearized.}

\begin{flalign} \label{eqn:surrogate_obj}
minimize \quad {1\over \vert S \vert }\sum_{i\in S} \vert w_{i,t_n}-w_{i,t_n-1}^m \vert+{max}_{i\in S}\vert w_{i,t_n}-w_{i,t_n-1}^m \vert &&
\end{flalign}

\subsubsection{Mixed Integer Linear Programming for Partial Replication}
To transform the nonlinear objective function (\ref{eqn:surrogate_obj}) into a linear one, we introduce dummy decision variables \(z_{i,t_n},Z_{t_n}\), convert (\ref{eqn:surrogate_obj}) into (\ref{eqn:surrogate_obj2}), and add constraints (\ref{eqn:constp1})--(\ref{eqn:constp3}) \cite{shanno1971linear,rardin1998optimization}. $z_{i,t_n}$ and $Z_{t_n}$ act like upper bounds for \( w_{i,t_n}-w_{i,t_n-1}^m \) and $z_{i,t_n}$, respectively.
% Because the objective function (\ref{eqn:surrogate_obj2}) is minimized, \(z_{i,t_n}\) and \(Z_{t_n}\) become \(\vert w_{i,t_n}-w_{i,t_n-1}^m \vert\) and \({max}_{i}~ z_{i,t_n}\), respectively, as the mathematical programming is optimized:
\PURPLE{Because the objective function (\ref{eqn:surrogate_obj2}) is minimized, \(z_{i,t_n}\) and \(Z_{t_n}\) become \(\vert w_{i,t_n}-w_{i,t_n-1}^m \vert\) and \({max}_{i}~ z_{i,t_n}\), respectively, as the mathematical programming is optimized:}

  \begin{flalign} \label{eqn:surrogate_obj2}
minimize \quad\quad {1\over \vert S \vert }\sum_{i\in S}z_{i,t_n}+Z_{t_n} &&
\end{flalign}

% $subject ~~ to $
 \begin{flalign*} 
 subject ~ to&& 
 \end{flalign*}
 \begin{align}
 \label{eqn:constp1} w_{i,t_n}-w_{i,t_n-1}^m &\leq z_{i,t_n} &, \forall i \in S\\
\label{eqn:constp2} w_{i,t_n}-w_{i,t_n-1}^m &\geq -z_{i,t_n} &, \forall i \in S\\
\label{eqn:constp3} z_{i,t_n} &\leq Z_{t_n} &, \forall i \in S\\
 \label{eqn:const1}\sum_{i\in S \backslash S^*} w_{i,t_n} \hat{\beta}_{i,t_n:t_n+T_A} &=  \overline{\beta}_{m,t_n:t_n+T_A} \\
 \label{eqn:const2}\sum_{i\in S \backslash S^*} w_{i,t_n} \hat{\alpha}_{i,t_n:t_n+T_A} &=  \overline{\alpha}_{m,t_n:t_n+T_A} \\
  \label{eqn:const3} \sum_{i\in S} w_{i,t_n} &= 1 \\
  \label{eqn:const4} w_{i,t_n} &\leq u_{i,t_n} &, \forall i \in S \\
  \label{eqn:const5} \sum_{i \in S} u_{i,t_n} &\leq N^* \\
  \label{eqn:const6} w_{i,t_n} &= w_{i,t_{n-1}}^m &, \forall i \in S^* \\
  \label{eqn:const7} 0 \leq w_{i,t_n} \leq 1  , ~~ u_{i,t_n}&\in \{0,1\}, z_{i,t_n}\geq 0, Z_{t_n} \geq 0 &, \forall i \in S
 \end{align}

Our portfolio construction scheme is expressed as a MILP model (\ref{eqn:surrogate_obj2})--(\ref{eqn:const7}) at each \(t_n\). 
\PURPLE{First, decision variables are listed in (\ref{eqn:const7}). }
Constraints (\ref{eqn:const1}) and (\ref{eqn:const2}) let us not predict the return of the market index \(m\) and minimize the errors as abovementioned. Because of the lack of time-series input data, such as newly listed stocks, \(\alpha_{i,t_n},\beta_{i,t_n}\) for some financial instruments \(S^*\) cannot be estimated at time \(t\). They are removed from the left-hand-side summations and right-hand-side calculations in (\ref{eqn:const1}) and (\ref{eqn:const2}). Instead, their weights are set to be their weights in the market index \(w_{i,t_n-1}^m\) as (\ref{eqn:const6}). 
Equality (\ref{eqn:const3}) enforces the sum of the weights to be one. 

We introduce the binary decision variable \(u_{i,t_n}\), indicating if a financial instrument \(i\) is included as defined in subsection \ref{subsec:overview_methods}. If a financial instrument \(i\) is in a portfolio, which means \(w_{i,t_n}>0\), to satisfy (\ref{eqn:const4}), \(u_{i,t_n}\) must be one. Inequality (\ref{eqn:const5}) is to \PURPLE{constrain} financial instruments not greater than a given constant \(N^* \in \mathbb{N}\) \cite{canakgoz2009mixed}, as explicated in section \ref{subsec:overview_methods}. Note that, because of some reasons (e.g. sinful stocks in stewardship codes), should a financial instrument $i$ have lower weight than some value $w_{i,t_n}^{max}$ or be excluded obligatorily, then an inclusion of a constraint $w_{i,t_n} \leq w_{i,t_n}^{max}~or ~w_{i,t_n}=0$ enables it. In the implementation, Python library PuLP is utilized to formulate and solve our MILP model.

\section{Experiments}
\label{section:exp}
% 작성원칙
% 결과 섹션에서, 여러분이 얻은 결과를 자세히 설명하기 위해서 과거 시제를 사용하세요
% 예: Overall, the expression of the AB gene had a positive correlation with the insertion of the XY gene.
% 그림, 표 및 그래프를 언급할 때는 현재 시제를 사용하세요.
% 예: Figure 1 displays the fast interlock interface signal flow.

\subsection{Input Data Structure}
  \begin{figure*}[pt]
  \centering
  \includegraphics[width=\linewidth]{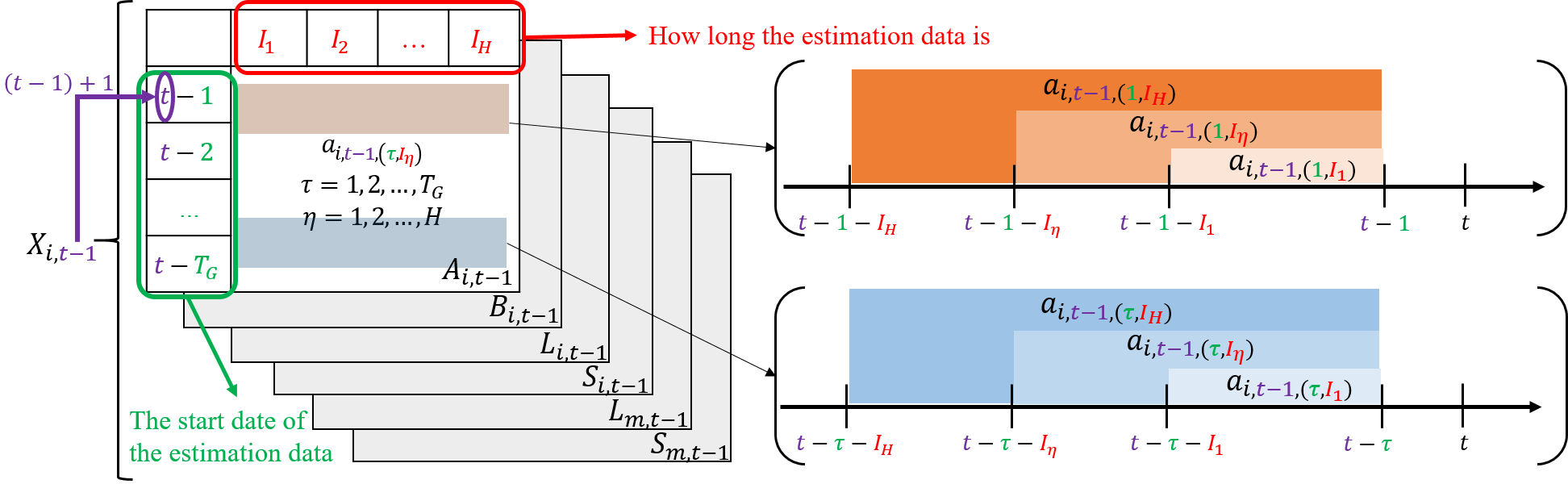}
  \caption{Architecture of Input Data $\textbf{X}_{i,t-1}$}
  % \Description{...}
  \label{fig:inp_data}
\end{figure*}

First, we introduce the input data structure to the DL model, depicted in Fig.~\ref{fig:inp_data}.
Input data for a financial instrument $i$ at trading date $t-1$ to predict $\alpha_{i,t:t+T_A}$, $\beta_{i,t:t+T_A}$, which is a constituent of market index $m$, is defined as a three-dimensional tensor $\textbf{X}_{i,t-1}=[A_{i,t-1},$ $B_{i,t-1},$ $L_{i,t-1},$ $S_{i,t-1},$ $L_{m,t-1},$ $S_{m,t-1}]\in \mathbb{R}^{K \times T_G \times H}$.
$K$ denotes the feature size; we use $K = 6$ features, and the shape of each is $T_G \times H$. Here,

\begin{itemize}
\item[$\bullet$] $A_{i,t-1}$: intercepts of the linearly regressed line with returns of $m$ and $i$ as the regressor and response variables, respectively;
\item[$\bullet$] $B_{i,t-1}$: slopes of the linearly regressed line with returns of $m$ and $i$ as the regressor and response variables, respectively;
\item[$\bullet$] $L_{i,t-1}$: averages of the excess returns of $i$;
\item[$\bullet$] $S_{i,t-1}$: standard deviations of the excess returns of $i$;
\item[$\bullet$] $L_{m,t-1}$: averages of the returns of market index $m$; and
\item[$\bullet$] $S_{m,t-1}$: standard deviations of the returns of market index $m$.
\end{itemize}

% $T_G$ and $H$ determine the start and end dates of the data for the estimation of the above six, respectively.
Take $A_{i,t-1}=[a_{i,t-1, (\tau,I_\eta)}]$ as an example. $a_{i,t-1, (\tau,I_\eta)}$ is the intercept of the linear regression line estimated in the data between $(t-1)+1-\tau-I_{\eta}$ and $(t-1)+1-\tau$. Row index $\tau$ determines the end dates of the data utilized to estimate the above statistics (intercepts, slopes, averages, and standard deviations). On the other hand, column $I_{\eta}$ determines how long the estimation data is, which is defined to capture the time-varying property of $\beta$.
% For instance
\ORANGE{That is}, when $\tau=1, \eta=1$ (the upper right in Fig.~\ref{fig:inp_data}),  $a_{i,t-1,(1,I_1)}$ is the intercept of the linear regression line estimated in the data between $t-1-I_1$ and $t-1$. Similarly, $l_{i,t-1,(1,I_1)}$, an element of $L_{i,t-1}$, is the average of the excess returns of $i$ from $t-1-I_1$ to $t-1$.

\subsection{Temporal Data Splitting To Preclude Look-Ahead Bias}
\label{subsec:data_splitting}

\begin{figure}[pt]
  \centering
  \includegraphics[width=0.95\linewidth]{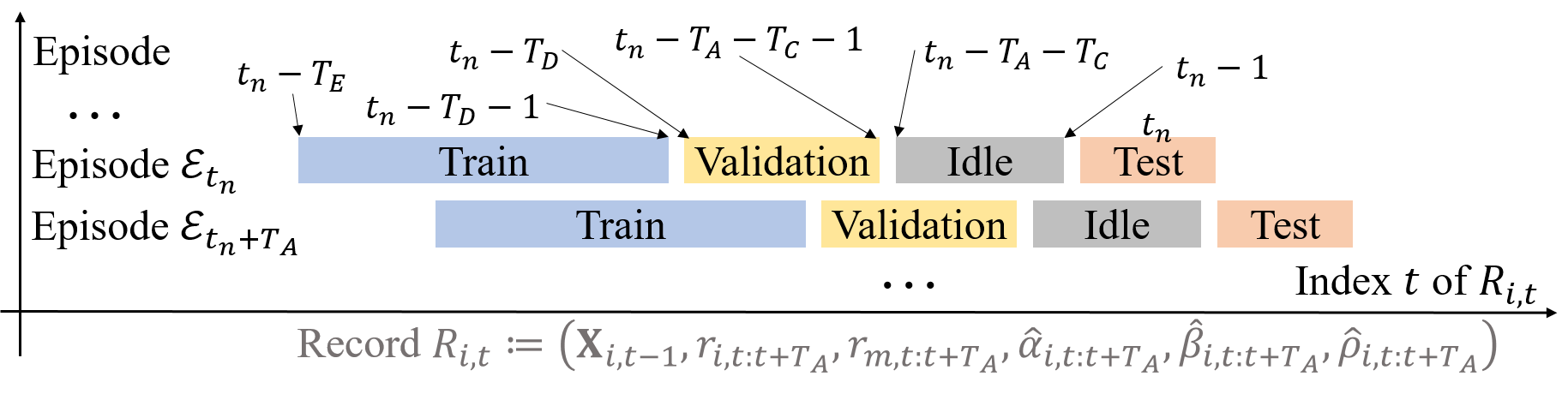}
  \caption{The Scheme of Experiments; Each row composed of the four blocks is an episode similar to the sliding window method \cite{hota2017time}. To prevent the look-ahead bias\cite{zhou2014active,isichenko2021quantitative}, an idle block is interposed.
  % , of which length is the sum of $T_A$ and $T_C$ required to calculate $r_{it},r_{mt}$ and $\alpha_{it},\beta_{it}$, respectively.
  }
  % \Description{...}
  \label{fig:exp_scheme}
\end{figure}

\PURPLE{
The whole data are} %split 
\BLUE{generated} 
\PURPLE{into episodes with period $T_A$ like the sliding window method \cite{hota2017time} to prevent %look-ahead bias \cite{zhou2014active,isichenko2021quantitative} 
shown in Fig.~\ref{fig:exp_scheme}. First, we define record $(i,t)$ as:}
\begin{itemize}
\item[$\bullet$] $R_{i,t}:=(\textbf{X}_{i,t-1}, \quad  r_{i,t:t+T_A}, r_{m,t:t+T_A}, \quad \tilde{\alpha}_{i,t:t+T_A},\tilde{\beta}_{i,t:t+T_A},\tilde{\rho}_{i,t:t+T_A})$
\end{itemize}
\PURPLE{
A record $R_{i,t}$ contains the data of input ($\textbf{X}_{i,t-1}$) and target ($\tilde{\alpha}_{i,t:t+T_A}$, $\tilde{\beta}_{i,t:t+T_A}$, $\tilde{\rho}_{i,t:t+T_A}$) for training $g^{[t_n]}$. 
It also contains $r_{i,t:t+T_A}$, $r_{m,t:t+T_A}$ to calculate the evaluation metrics $\hat{PE}, \hat{TE}$. 
Next, we define each block and episode as:
}
% \begin{itemize}
% \item[\bullet] Train block $t_n$: $B_{i,t_n}^{Train}:=\{R_{i,t}: t=t_n-T_E,...,t_n-T_D-1\}$
% \item[\bullet] Validation block $t_n$:  $B^{Vali}_{i,t_n}:=\{R_{i,t}: t=t_n-T_D,...,t_n-T_A-T_C-1\}$
% \item[\bullet] Idle block $t_n$: $B_{i,t_n}^{Idle}:=\{R_{i,t}: t=t_n-T_A-T_C,...,t_n-1\}$
% \item[\bullet] Test block $t_n$: $B^{Test}_{i,t_n}:=\{R_{i,t_n}\}$
% \item[\bullet] Episode $t_n$: $\mathcal{E}_{i,t_n}:=\{B_{i,t_n}^{Train},B_{i,t_n}^{Vali},B_{i,t_n}^{Idle},B_{i,t_n}^{Test}\}$
% \end{itemize}
\begin{itemize}
\item[$\bullet$] Train block $t_n$: $B_{i,t_n}^{Train}:=\{R_{i,t}: t \in [t_n-T_E,~t_n-T_D-1] \cap \mathbb{Z}\}$
\item[$\bullet$] Validation block $t_n$:  $B^{Vali}_{i,t_n}:=\{R_{i,t}: t \in [t_n-T_D,~ t_n-T_A-T_C-1] \cap \mathbb{Z}\}$
\item[$\bullet$] Idle block $t_n$: $B_{i,t_n}^{Idle}:=\{R_{i,t}: t \in [t_n-T_A-T_C,~t_n-1] \cap \mathbb{Z}\}$
\item[$\bullet$] Test block $t_n$: $B^{Test}_{i,t_n}:=\{R_{i,t}: t= t_n\}$
\item[$\bullet$] Episode $t_n$: $\mathcal{E}_{i,t_n}:=\{B_{i,t_n}^{Train},B_{i,t_n}^{Vali},B_{i,t_n}^{Idle},B_{i,t_n}^{Test}\}$
\end{itemize}

The objective for defining $\mathcal{E}_{i,t_n}$ is to provide adequate data for two tasks: (a) training and evaluating $g^{[t_n]}$, and (b) generating and evaluating $\textbf{w}_{t_n}$. 
First, $g^{[t_n]}$ is trained based on $B_{i,t_n}^{Train}$ and $B^{Vali}_{i,t_n}$. % the train and validation blocks in $\mathcal{E}_{i,t_n}$. 
Next, $g^{[t_n]}$ and $\textbf{w}_{t_n}$ are evaluated with $B^{Test}_{i,t_n}$. 
As explained in subsection \ref{subsec:overview_methods}, the purpose for training $g^{[t_n]}$ is to predict $\hat{\alpha}_{i,t_n:t_n+T_A}$, $\hat{\beta}_{i,t_n:t_n+T_A}$ given input $\textbf{X}_{i, t_n-1}$, and $\textbf{w}_{t_n}$ is generated by our MILP model utilizing $\hat{\alpha}_{i,t_n:t_n+T_A}$, $\hat{\beta}_{i,t_n:t_n+T_A}$, $\textbf{X}_{i, t_n-1}$. 
This can be evaluated by $R_{i,t_n}$. 
Thus, $B^{Test}$ consists of only one record $R_{i,t_n}$. 
% the test block of $\mathcal{E}_{i,t_n}$ consists of only one record $R_{i,t_n}$ that contains $\tilde{\alpha}_{i,t_n:t_n+T_A}$, $\tilde{\beta}_{i,t_n:t_n+T_A}, \textbf{X}_{i, t_n-1}$.
% In addition, $r_{i,t_n:t_n+t_A}$ is utilized to evaluate $\textbf{w}_{t_n}$ generated by our MILP model that utilizes $\hat{\alpha}_{i,t_n:t_n+T_A}$, $\hat{\beta}_{i,t_n:t_n+T_A}$, $\textbf{X}_{i, t_n-1}$.
 
% Because $\mathcal{E}_{i,t^*}$'s index is defined to be the same as the index of the record of its test block, 
% The train and validation blocks are employed in the training process of $g^{[t_n]}$. 
% The test block is utilized for $g^{[t_n]}$ to 
% $\mathcal{E}_{i,t_n}$ means that 

\BLUE{
Unlike usual train-validation-test splits, the idle block, $B_{i,t_n}^{Idle}$, is interposed between $B_{i,t_n}^{Train}$ and $B^{Vali}_{i,t_n}$ and not employed for (a) and (b) since utilizing $B_{i,t_n}^{Idle}$ can cause a look-ahead bias \cite{zhou2014active,isichenko2021quantitative}. 
% The corresponding train and validation blocks are employed to train and select $g^{[t_n]}$, respectively. 
} 
% \PURPLE{Unlike usual train-validation-test splits, we interpose the idle blocks between the validation and test blocks because utilizing the data of idle block causes a look-ahead bias \cite{zhou2014active,isichenko2021quantitative}. }
Consider $R_{i,t_n-T_A-T_C}$, the first record of $B_{i,t_n}^{Idle}$, as an example. It contains $\tilde{\beta}_{i,t_n-T_A-T_C:t_n-T_C}$ by its definition.
% Because we use $r_{i,t_n-T_A:t_n}$ to estimate $\tilde{\beta}_{i,t_n-T_A-T_C:t_n-T_C}$ as explained in subsection \ref{subsec:sl}, the first record of the idle block utilizes $p_{i, t_n}$.} 
\BLUE{
As explained in subsection \ref{subsec:sl}, $r_{i,t_n-T_A:t_n}$ is required to estimate $\tilde{\beta}_{i,t_n-T_A-T_C:t_n-T_C}$. So, $p_{i, t_n}$ is required for $R_{i,t_n-T_A-T_C}$ to compute $r_{i,t_n-T_A:t_n}$ by (\ref{eqn:def_ret}). 
However, when generating $\textbf{w}_{t_n}$ and training $g^{[t_n]}$, $p_{i, t_n}$ should not be utilized because we should not utilize the data at $t_n$, as explained in subsection \ref{subsec:overview_methods}. That is, $R_{i,t_n-T_A-T_C} \subset B_{i,t_n}^{Idle}$ should not be used. Therefore, $B_{i,t_n}^{Idle}$ is interposed and not utilized.
% However, $p_{i, t_n}$ is unavailable when generating $\textbf{w}_{t_n}$ because we cannot build a portfolio weighted by $\textbf{w}_{t_n}$ with data \textbf{at} $t_n$, as explained in subsection \ref{subsec:overview_methods}. 
% % (We can generate $\textbf{w}_{t_n}$ with data \textbf{before} $t_n$).
% In addition, $g^{[t_n]}$ must be trained before generating $\textbf{w}_{t_n}$ because the predicted values of $g^{[t_n]}$ are used for generating $\textbf{w}_{t_n}$. 
% That is, $g^{[t_n]}$ cannot be trained with $\tilde{\beta}_{i,t_n-T_A-T_C:t_n-T_C}$ which is unavailable before generating $\textbf{w}_{t_n}$.
% Thus, should $\tilde{\beta}_{i,t_n-T_A-T_C:t_n-T_C} \subset R_{i,t_n-T_A-T_C}$ be utilized for training $g^{[t_n]}$, the experiment has a look-ahead bias. 
% Hence, $B_{i,t_n}^{Idle}$ is interposed and not utilized.
}

\subsection{Experimental Settings}

We utilized data on the daily price return of KOSPI 200 and its constituents, as well as the weight of each constituent. All of them from January 2000 to June 2022 were acquired from the Korea Stock Exchange. \PURPLE{Python and PyTorch were utilized for the data processing, implementation, and experiments.}

We employed the canonical neural networks as the feature extractor $\phi^{[t_n]}$ in Fig.~\ref{fig:dl_arc}: MLP, LSTM, GRU, and Transformer. CDF-transformed $\textbf{X}_{i,t-1}$ is transposed or flattened to be appropriate for each feature extractor. 
\PURPLE{The major hyperparameters for the four types of $\phi^{[t_n]}$ were set to be as follows: \textit{hidden\_size} = \textit{dim\_feedforward} = 64, \textit{num\_layers} = 2, and \textit{dropout} = 0.1.
% \begin{itemize}
% \item[\bullet]   \textit{hidden\_size} = \textit{dim\_feedforward} = 64
% \item[\bullet]  \textit{num\_layers} = 2 
% \item[\bullet]  \textit{dropout} = 0.1
% \end{itemize} 
The FC $\psi_{\bullet}^{[t_n]}$ in Fig.~\ref{fig:dl_arc} consists of one hidden layer and one output layer, whose activation functions were Leaky ReLU and a sigmoid, respectively, with dropout layers.} 

\begin{table}[]
  \caption{Yearly Performance ($\hat{PE}$) of Historical Estimations for Various Periods' Data for Estimation}
  \label{table:hist_estimation}
  \begin{center}
\begin{tabular}{p{0.15\linewidth}p{0.09\linewidth}p{0.09\linewidth}p{0.09\linewidth}p{0.09\linewidth}p{0.09\linewidth}p{0.09\linewidth}}
\toprule
\textbf{Year\textbackslash{}Days} & \textbf{504}    & \textbf{756}    & \textbf{1,008}   & \textbf{1,260}   & \textbf{1,512}   & \textbf{2,520}   \\ \midrule
\textbf{2016}                     & 0.0195          & 0.0162          & 0.0155          & 0.0152          & \textbf{0.0152} & 0.0168          \\
\textbf{2017}                     & 0.0148          & 0.0148          & 0.0139          & \textbf{0.0136} & 0.0136          & 0.0149          \\
\textbf{2018}                     & \textbf{0.0184} & 0.0192          & 0.0199          & 0.0196          & 0.0195          & 0.0211          \\
\textbf{2019}                     & 0.0161          & \textbf{0.0144} & 0.0145          & 0.0151          & 0.0149          & 0.0152          \\
\textbf{2020}                     & 0.0261          & 0.0255          & \textbf{0.0251} & 0.0253          & 0.0253          & 0.0253          \\
\textbf{2021}                     & 0.0237          & 0.0232          & 0.0229          & 0.0221          & 0.0219          & \textbf{0.0213} \\
\textbf{2022}                     & 0.0230          & 0.0211          & 0.0211          & 0.0205          & 0.0202          & \textbf{0.0199} \\ \bottomrule
\end{tabular}
\end{center}
\end{table}

The prediction models are optimized by the batch gradient descent method with a batch-size of 512, momentum of 0.1, and L2 regularization of 1.0e-04. 
Cosine annealing is employed as a learning rate scheduler with the initial learning rate of 1.0e-02 and max epochs of 100. 
Early stopping is also applied. 
For the loss function, mean squared error is chosen. 
Additionally, $T_C$ is set to be two.

% We utilize PE (\ref{eqn:PE_est}) and TE (\ref{eqn:TE_est}) to evaluate the performance of the predictions of $\hat{\alpha}_{it},\hat{\beta}_{it}$ and the portfolio construction for a given set of time steps $\mathfrak{T}$ and $V=\{(i,t) : i \in S ~~and~~ t \in \mathfrak{T} \}$. These equations are derived from (\ref{eqn:obj_pred_} and \ref{eqn:obj_port_}) by replacing $\mathbb{E}$ with the average. 
% \begin{align}
% % \label{eqn:PE_est}
% % &\hat{PE}={1 \over |V|} \sum_{v\in V} (r_{it} - \hat{\beta}_{it} r_{mt} -\hat{\alpha}_{it})^2 \\
% \label{eqn:TE_est}
% &\hat{TE}(r_\pi,r_m) = {1 \over \vert \mathfrak{T} \vert} \sum_{t\in \mathfrak{T}} (r_{\pi t}-r_{mt})^2
% \end{align}

\subsection{Evaluation of the Historical Estimation}

We first evaluate the performance ($\hat{PE}$) of historical estimations, defined in subsection~\ref{sub:related_betas}, for $\alpha_{i,t_n:t_n+T_A}$, $\beta_{i,t_n:t_n+T_A}$ in Table \ref{table:hist_estimation}. The table shows that the best data period for estimation varies as time passes. Because asset managers cannot know which estimation period will be the best, they need a systematic method to integrate these historical estimations into one. This necessity can be fulfilled by our prediction method which integrates the multi-period historical data and outputs a single pair of $\hat{\alpha}_{i,t_n:t_n+T_A}$, $\hat{\beta}_{i,t_n:t_n+T_A}$. 
Note that various estimation periods are tested, and the best ones in each year are selected to be shown in Table \ref{table:hist_estimation}.

Another interesting result in Table \ref{table:hist_estimation} is that the worst performance across all estimation periods is observed in 2020, the year when the COVID-19 pandemic remarkably struck the global economy. The reason for this is likely the unprecedented government restrictions on society that year, as suggested by \cite{baker2020unprecedented}.

% \usepackage{booktabs}
\begin{table}[pt]
  \caption{Yearly Performances ($\hat{PE}$) of the Deep-Learning Models and Best Historical Estimation}
  \label{tab:sl_results}
  \begin{center}
\begin{tabular}{p{0.11\columnwidth}p{0.11\columnwidth}p{0.11\columnwidth}p{0.11\columnwidth}p{0.11\columnwidth}p{0.11\columnwidth}}
\toprule
\textbf{Year}    & \multicolumn{1}{c}{\textbf{GRU}} & \multicolumn{1}{c}{\textbf{LSTM}} & \multicolumn{1}{c}{\textbf{Trans}} & \multicolumn{1}{c}{\textbf{MLP}} & \multicolumn{1}{c}{\textbf{Historical}} \\ \midrule
\textbf{2016}    & 0.00708                         & 0.00690                          & 0.00780                           & \textbf{0.00683}                         & 0.01519                                \\
\textbf{2017}    & 0.00892                         & \textbf{0.00890}                          & 0.00917                           & 0.00896                         & 0.01362                                \\
\textbf{2018}    & 0.00919                         & \textbf{0.00886}                          & 0.00941                           & 0.00891                         & 0.01845                                \\
\textbf{2019}    & 0.00615                         & 0.00596                          & 0.00673                           & \textbf{0.00592}                         & 0.01435                                \\
\textbf{2020}    & 0.01791                         & 0.01785                          & \textbf{0.01776}                           & 0.01792                         & 0.02511                                \\
\textbf{2021}    & 0.01244                         & \textbf{0.01223}                          & 0.01329                           & 0.01237                         & 0.02126                                \\
\textbf{2022}    & 0.01084                         & 0.01038                          & 0.01105                           & \textbf{0.01033}                         & 0.01988                                \\ \midrule
\textbf{Average} & 0.01036                         & \textbf{0.01015}                          & 0.01075                           & 0.01018                         & 0.01826                                \\ \bottomrule

\end{tabular}
\end{center}
\end{table}

\begin{table}[pt]
  \caption{Tracking Errors Using Our Method and Full Replication  (Unit: 1.0e-5)}
  \label{tab:prot_results}
\begin{center}
\begin{tabular}{cccccc}
\toprule
\textbf{\begin{tabular}[c]{@{}c@{}}The Number\\ of Stocks\end{tabular}} & \textbf{MLP}   & \textbf{LSTM}  & \textbf{GRU}   & \textbf{Trans} & \textbf{Full} \\ \midrule
\textbf{30}                                                             & 7.548          & 7.522          & 8.415          & 7.082          & 1.797         \\
\textbf{40}                                                             & 4.701          & 4.259          & 4.570          & 3.522          & 1.797         \\
\textbf{50}                                                             & 2.951          & 3.013          & 3.153          & 2.995          & 1.797         \\
\textbf{60}                                                             & 2.383          & 2.170          & 2.257          & 2.307          & 1.797         \\
\textbf{70}                                                             & 2.042          & 2.095          & 1.947          & \textbf{1.778} & 1.797         \\
\textbf{80}                                                             & 1.861          & \textbf{1.797} & 1.867          & 1.836          & 1.797         \\
\textbf{90}                                                             & \textbf{1.741} & \textbf{1.768} & 1.831          & 1.812          & 1.797         \\
\textbf{100}                                                            & \textbf{1.747} & \textbf{1.664} & 1.797          & \textbf{1.782} & 1.797         \\
\textbf{110}                                                            & \textbf{1.770} & \textbf{1.764} & \textbf{1.787} & \textbf{1.732} & 1.797         \\
\textbf{120}                                                            & 1.835          & \textbf{1.745} & \textbf{1.794} & 1.821          & 1.797         \\
\textbf{130}                                                            & \textbf{1.756} & \textbf{1.680} & \textbf{1.733} & \textbf{1.726} & 1.797         \\
\textbf{140}                                                            & \textbf{1.727} & \textbf{1.751} & \textbf{1.776} & \textbf{1.776} & 1.797         \\
\textbf{150}                                                            & \textbf{1.738} & \textbf{1.700} & \textbf{1.722} & 1.820          & 1.797         \\
\textbf{160}                                                            & \textbf{1.740} & \textbf{1.794} & \textbf{1.779} & 1.809          & 1.797         \\
\textbf{170}                                                            & \textbf{1.766} & 1.812          & \textbf{1.782} & 1.804          & 1.797         \\
\textbf{180}                                                            & \textbf{1.763} & \textbf{1.762} & \textbf{1.769} & 1.801          & 1.797         \\
\textbf{190}                                                            & \textbf{1.774} & \textbf{1.788} & \textbf{1.788} & 1.807          & 1.797         \\ \bottomrule
\end{tabular}
\end{center}
\end{table}
\subsection{Prediction of Market Sensitivities and Alphas}

Now, we compare the performances of our models with the historical estimations in Table \ref{tab:sl_results}. The table shows that all our models perform better than the best performance of the historical estimations (Column \textbf{Historical}) in all years. Moreover, the performance improvement is significant. The average $\hat{PE}$ of our methods is around 57$\%$ of that of the best historical estimation, indicating that our models are effective.

In Table \ref{tab:sl_results}, we observe that our models also performed the worst in 2020, which is similar to the performance of historical estimations in the same year. We may conjecture that our models were vulnerable to the unprecedented government restrictions  \cite{baker2020unprecedented}. So, developing a methodology to reflect them would be a future research topic to improve our models. %Moreover
\BLUE{In addition}, the best models are different for each year in Table \ref{tab:sl_results}. To integrate them into a single model, applying ensemble techniques  \cite{dong2020survey,ganaie2021ensemble} would be future work. Moreover, we utilize a single-factor model, which can be extended to multi-factor models like using the factors defined by \cite{fama2015five}.

\subsection{Performance of the Portfolios}
So far, we have shown that all our prediction models outperform the historical estimations. In this subsection, we evaluate our portfolio construction method that uses %our predictions on  $\hat{\alpha}_{i,t_n:t_n+T_A}, \hat{\beta}_{i,t_n:t_n+T_A}$.
$\hat{\alpha}_{i,t_n:t_n+T_A}, \hat{\beta}_{i,t_n:t_n+T_A}$ predicted by our prediction models.

\begin{figure}[pt]
  \centering
  \includegraphics[width=0.82\linewidth]{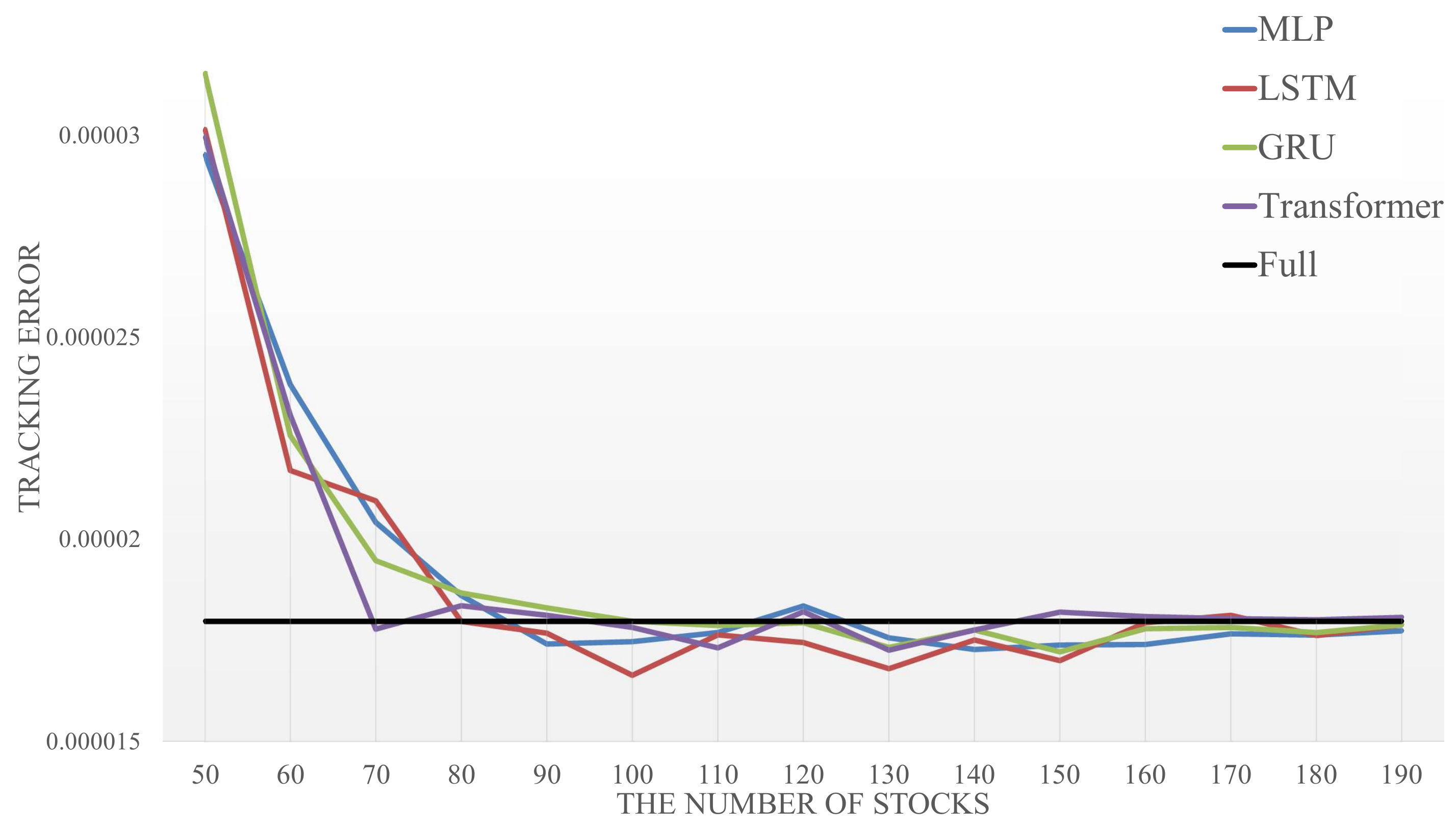}
  \caption{Comparison of Tracking Errors; The changes of $\hat{TE}$ as the limitations on the number of stocks $N^*$ changes.
  % MLP, LSTM, GRU, Transformer, and Full are  multi-layer perceptrons, long-short term memory, gated-recurrent unit, and the encoder of Transformer, respectively.
  }
  % \Description{The changes of $\hat{TE}$  as the limitations on the number of stocks $N^*$ changes}
  \label{fig:port_perform}
\end{figure}

To evaluate our method, we compare the $\hat{TE}$ of our method and the full replications labeled as \textbf{Full} in Table \ref{tab:prot_results} and Fig.~\ref{fig:port_perform}. 
The $\hat{TE}$ of full replication is measured under the condition that the portfolio is updated according to the market-capitalization ratio one day before the update because subsection \ref{subsec:overview_methods} assumes that data up to one day before the execution date are available. 
\BLUE{That is, when we implement full replication, it cannot exactly track the constituents and weights of the market index in real-time, so its $\hat{TE}$ cannot be ideal zero in Table \ref{tab:prot_results}.} 
Since the full replication is irrelevant to the number of constituents, tracking error is represented as just one value as shown as "Full" in Table \ref{tab:prot_results} and Fig.~\ref{fig:port_perform}.

When $N^*\geq90$, $\hat{TE}$  of our methods are similar to those of the full replication in Fig.~\ref{fig:port_perform}, which implies that the 90 stocks selected and weighted by our method are enough to replicate the given KOSPI 200. 
Note that the number 90 is less than half of the constituents of KOSPI 200. 
In Table \ref{tab:prot_results}, 75$\%$ of our methods (the bold numbers) showed better tracking errors than the full replications, when $N^*\geq90$. Therefore, we have shown that our portfolio construction methodology with the predicted market sensitivities is effective.

Additionally, we surmise that the reason why the performance decreased after 90 stocks is the idiosyncratic errors $\epsilon_{i,t_n:t_n+T_A}$ cannot be offset sufficiently with the fewer number of stocks. Hence, taking into account the correlation between the errors may allow us to lower our lower bound of 90 stocks; this would be future work. Also, the fewer number of stocks may have increased the difference in the exposure to other factors between the portfolios and market index. This might be solved by considering multi-factor models like \cite{fama2015five}, adding new outputs for the new factors to the prediction models, and introducing constraints that make the net sensitivities of them be the those of the market. Despite the room for future works, the replication of a market index with fewer than the half number of its constituents is significant because it can reduce the management costs of funds with decreasing the number of stocks. Moreover, our method can flexibly reflect other constraints occurring in actual deployment.

\section{Conclusion}
\label{section:conclusion}

This study has proposed a two-step novel approach to the partial replication of a market index. 
In the first step, we examine the use of several deep-learning models to predict the market sensitivities of index constituents. 
Amid the prediction for market sensitivities, we present 
% the applications of practical data handling methods: 
\PURPLE{CDF transformation and a target-data-generation method}.
Then, we design a mixed-integer linear programming model to construct an index\BLUE{-tracking} portfolio, \BLUE{which uses} the predicted market sensitivities in the first step. Experimental results in KOSPI 200 indicated that our prediction models had only 57$\%$ of the errors of historical estimations. 
Also, half the number of the whole constituents was enough for our portfolio construction method to mimic KOSPI 200. 
To our knowledge, this is the first study to demonstrate the efficacy of deep-learning architectures in predicting market sensitivities with the pragmatic partial-index-tracking method that controls the predicted market sensitivities.

% \input{text/Example}

% \bibliographystyle{plain}
\bibliographystyle{splncs03_unsrt.bst}
\bibliography{reference.bib}

\begin{thebibliography}{10}
\providecommand{\url}[1]{\texttt{#1}}
\providecommand{\urlprefix}{URL }

\bibitem{heath2022index}
Heath, D., Macciocchi, D., Michaely, R., Ringgenberg, M.C.: Do index funds
  monitor? The Review of Financial Studies  35(1),  91--131 (2022)

\bibitem{oh2005using}
Oh, K.J., Kim, T.Y., Min, S.: Using genetic algorithm to support portfolio
  optimization for index fund management. Expert Systems with applications
  28(2),  371--379 (2005)

\bibitem{kim2020index}
Kim, S., Kim, S.: Index tracking through deep latent representation learning.
  Quantitative Finance  20(4),  639--652 (2020)

\bibitem{chang2004evaluating}
Chang, K.P.: Evaluating mutual fund performance: an application of minimum
  convex input requirement set approach. Computers \& Operations Research
  31(6),  929--940 (2004)

\bibitem{keim1999analysis}
Keim, D.B.: An analysis of mutual fund design: the case of investing in
  small-cap stocks. Journal of Financial Economics  51(2),  173--194 (1999)

\bibitem{sharpe1972risk}
Sharpe, W.F.: Risk, market sensitivity and diversification. Financial Analysts
  Journal  28(1),  74--79 (1972)

\bibitem{sharpe1995risk}
Sharpe, W.F.: Risk, market sensitivity, and diversification. Financial Analysts
  Journal  51(1),  84--88 (1995)

\bibitem{blume1975timevarying}
Blume, M.E.: Betas and their regression tendencies. The Journal of Finance
  30(3),  785--795 (1975),
  \url{https://onlinelibrary.wiley.com/doi/abs/10.1111/j.1540-6261.1975.tb01850.x}

\bibitem{ferson1993risk}
Ferson, W.E., Harvey, C.R.: The risk and predictability of international equity
  returns. Review of financial Studies  6(3),  527--566 (1993)

\bibitem{pagan1980kalman}
Pagan, A.: Some identification and estimation results for regression models
  with stochastically varying coefficients. Journal of Econometrics  13(3),
  341--363 (1980),
  \url{https://www.sciencedirect.com/science/article/pii/0304407680900846}

\bibitem{das2010market}
Das, A., Ghoshal, T.K.: Market risk beta estimation using adaptive kalman
  filter. International Journal of Engineering Science and Technology  2(6),
  1923--1934 (2010)

\bibitem{engle2002dcc}
Engle, R.: Dynamic conditional correlation: A simple class of multivariate
  generalized autoregressive conditional heteroskedasticity models. Journal of
  Business \& Economic Statistics  20(3),  339--350 (2002),
  \url{http://www.jstor.org/stable/1392121}

\bibitem{engle2016garch}
Engle, R.F.: {Dynamic Conditional Beta}. Journal of Financial Econometrics
  14(4),  643--667 (08 2016), \url{https://doi.org/10.1093/jjfinec/nbw006}

\bibitem{siegel1995option}
Siegel, A.F.: Measuring systematic risk using implicit beta. Management Science
   41(1),  124--128 (1995), \url{http://www.jstor.org/stable/2632906}

\bibitem{buss2012option}
Buss, A., Vilkov, G.: {Measuring Equity Risk with Option-implied Correlations}.
  The Review of Financial Studies  25(10),  3113--3140 (08 2012),
  \url{https://doi.org/10.1093/rfs/hhs087}

\bibitem{hollstein2016beta}
Hollstein, F., Prokopczuk, M.: Estimating beta. Journal of Financial and
  Quantitative Analysis  51(4),  1437–1466 (2016)

\bibitem{faff2000ts}
Faff, R.W., Hillier, D., Hillier, J.: Time varying beta risk: An analysis of
  alternative modelling techniques. Journal of Business Finance \& Accounting
  27(5-6),  523--554 (2000),
  \url{https://onlinelibrary.wiley.com/doi/abs/10.1111/1468-5957.00324}

\bibitem{hota2017time}
Hota, H., Handa, R., Shrivas, A.: Time series data prediction using sliding
  window based rbf neural network. International Journal of Computational
  Intelligence Research  13(5),  1145--1156 (2017)

\bibitem{hopfield1984neurons}
Hopfield, J.J.: Neurons with graded response have collective computational
  properties like those of two-state neurons. Proceedings of the national
  academy of sciences  81(10),  3088--3092 (1984)

\bibitem{fernandez2007portfolio}
Fern{\'a}ndez, A., G{\'o}mez, S.: Portfolio selection using neural networks.
  Computers \& operations research  34(4),  1177--1191 (2007)

\bibitem{kwak2021neural}
Kwak, Y., Song, J., Lee, H.: Neural network with fixed noise for index-tracking
  portfolio optimization. Expert Systems with Applications  183,  115298 (2021)

\bibitem{ouyang2019index}
Ouyang, H., Zhang, X., Yan, H.: Index tracking based on deep neural network.
  Cognitive Systems Research  57,  107--114 (2019)

\bibitem{bradrania2022state}
Bradrania, R., Pirayesh~Neghab, D., Shafizadeh, M.: State-dependent stock
  selection in index tracking: a machine learning approach. Financial Markets
  and Portfolio Management  36(1),  1--28 (2022)

\bibitem{zhengchen2020}
Zheng, Y., Chen, B., Hospedales, T.M., Yang, Y.: Index tracking with
  cardinality constraints: A stochastic neural networks approach. Proceedings
  of the AAAI Conference on Artificial Intelligence  34(01),  1242--1249 (Apr
  2020), \url{https://ojs.aaai.org/index.php/AAAI/article/view/5478}

\bibitem{wolfgang2022ml}
Drobetz, W., Hollstein, F., Otto, T., Prokopczuk, M.: Estimating stock market
  betas via machine learning (2021), available at SSRN:
  https://ssrn.com/abstract=3933048

\bibitem{chang2011option}
Chang, B.Y., Christoffersen, P., Jacobs, K., Vainberg, G.: {Option-Implied
  Measures of Equity Risk*}. Review of Finance  16(2),  385--428 (03 2011),
  \url{https://doi.org/10.1093/rof/rfq029}

\bibitem{sr2005option}
Skintzi, V.D., Refenes, A.P.N.: Implied correlation index: A new measure of
  diversification. Journal of Futures Markets  25(2),  171--197 (2005),
  \url{https://onlinelibrary.wiley.com/doi/abs/10.1002/fut.20137}

\bibitem{kks2014option}
Kempf, A., Korn, O., Saßning, S.: {Portfolio Optimization Using
  Forward-Looking Information*}. Review of Finance  19(1),  467--490 (03 2014),
  \url{https://doi.org/10.1093/rof/rfu006}

\bibitem{luenberger2009investment}
Luenberger, D., et~al.: Investment science: International edition. OUP
  Catalogue  (2009)

\bibitem{zhang2018definition}
Zhang, W., Yang, G., Lin, Y., Ji, C., Gupta, M.M.: On definition of deep
  learning. In: 2018 World automation congress (WAC). pp. 1--5. IEEE (2018)

\bibitem{hochreiter1997long}
Hochreiter, S., Schmidhuber, J.: Long short-term memory. Neural computation
  9(8),  1735--1780 (1997)

\bibitem{cho2014learning}
Cho, K., Van~Merri{\"e}nboer, B., Gulcehre, C., Bahdanau, D., Bougares, F.,
  Schwenk, H., Bengio, Y.: Learning phrase representations using rnn
  encoder-decoder for statistical machine translation. arXiv preprint
  arXiv:1406.1078  (2014)

\bibitem{vaswani2017attention}
Vaswani, A., Shazeer, N., Parmar, N., Uszkoreit, J., Jones, L., Gomez, A.N.,
  Kaiser, {\L}., Polosukhin, I.: Attention is all you need. Advances in neural
  information processing systems  30 (2017)

\bibitem{gardner1998artificial}
Gardner, M.W., Dorling, S.: Artificial neural networks (the multilayer
  perceptron)—a review of applications in the atmospheric sciences.
  Atmospheric environment  32(14-15),  2627--2636 (1998)

\bibitem{fu2016using}
Fu, R., Zhang, Z., Li, L.: Using lstm and gru neural network methods for
  traffic flow prediction. In: 2016 31st Youth Academic Annual Conference of
  Chinese Association of Automation (YAC). pp. 324--328. IEEE (2016)

\bibitem{zhou2014active}
Zhou, X., Jain, S.: Active equity management (2014)

\bibitem{isichenko2021quantitative}
Isichenko, M.: Quantitative Portfolio Management: The Art and Science of
  Statistical Arbitrage. Wiley (2021),
  \url{https://books.google.co.kr/books?id=lHcxzgEACAAJ}

\bibitem{pope1994discovering}
Pope, P.F., Yadav, P.K.: Discovering errors in tracking error. Journal of
  Portfolio Management  20(2),  27--32 (1994)

\bibitem{roll1992mean}
Roll, R.: A mean/variance analysis of tracking error. Journal of portfolio
  management  18(4),  13--22 (1992)

\bibitem{canakgoz2009mixed}
Canakgoz, N.A., Beasley, J.E.: Mixed-integer programming approaches for index
  tracking and enhanced indexation. European Journal of Operational Research
  196(1),  384--399 (2009)

\bibitem{casella2021statistical}
Casella, G., Berger, R.L.: Statistical inference. Cengage Learning (2021)

\bibitem{wang2005asymptotics}
Wang, X.: Asymptotics of the theil--sen estimator in the simple linear
  regression model with a random covariate. Journal of Nonparametric Statistics
   17(1),  107--120 (2005)

\bibitem{dang2008theil}
Dang, X., Peng, H., Wang, X., Zhang, H.: Theil-sen estimators in a multiple
  linear regression model. Olemiss Edu  (2008)

\bibitem{pafka2003noisy}
Pafka, S., Kondor, I.: Noisy covariance matrices and portfolio optimization ii.
  Physica A: Statistical Mechanics and its Applications  319,  487--494 (2003)

\bibitem{shanno1971linear}
Shanno, D.F., Weil, R.L.: “linear” programming with absolute-value
  functionals. Operations Research  19(1),  120--124 (1971)

\bibitem{rardin1998optimization}
Rardin, R.L., Rardin, R.L.: Optimization in operations research, vol. 166.
  Prentice Hall Upper Saddle River, NJ (1998)

\bibitem{baker2020unprecedented}
Baker, S.R., Bloom, N., Davis, S.J., Kost, K.J., Sammon, M.C., Viratyosin, T.:
  The unprecedented stock market impact of covid-19. Tech. rep., national
  Bureau of economic research (2020)

\bibitem{dong2020survey}
Dong, X., Yu, Z., Cao, W., Shi, Y., Ma, Q.: A survey on ensemble learning.
  Frontiers of Computer Science  14(2),  241--258 (2020)

\bibitem{ganaie2021ensemble}
Ganaie, M.A., Hu, M., et~al.: Ensemble deep learning: A review. arXiv preprint
  arXiv:2104.02395  (2021)

\bibitem{fama2015five}
Fama, E.F., French, K.R.: A five-factor asset pricing model. Journal of
  financial economics  116(1),  1--22 (2015)

\end{thebibliography}

\end{document}